\documentclass[preprint,showpacs,preprintnumbers,amsmath,amssymb]{revtex4-1}

\usepackage{graphicx}
\usepackage{dcolumn}
\usepackage{bm}
\usepackage{amsmath,empheq}
\usepackage{natbib}

\begin{document}
\title{Dynamics of impurities in a three-dimensional volume-preserving map}
\author{Swetamber Das}
\email{swetdas@physics.iitm.ac.in}
\affiliation{Department of Physics, Indian Institute of Technology Madras, Chennai, 600036, India.}
\author{Neelima Gupte}
\email{gupte@physics.iitm.ac.in}
\affiliation{Department of Physics, Indian Institute of Technology Madras, Chennai, 600036, India.}
\begin{abstract}
We study the dynamics of inertial particles in three dimensional incompressible maps, as representations of volume preserving flows. The impurity dynamics has been modeled, in the Lagrangian framework, by a six-dimensional dissipative bailout embedding map. The fluid-parcel dynamics of the base map is embedded in the particle dynamics governed by the map. The base map considered for the present study is the Arnold-Beltrami-Childress (ABC) map. We consider the  behavior of the system both in the aerosol regime, where the density of the particle is larger than that of the base flow, as well as the bubble regime, where the particle density is less than that of the base flow. The phase spaces in both the regimes show rich and complex dynamics with three type of dynamical behaviors - chaotic structures, regular orbits and hyperchaotic regions. In the one-action case, the aerosol regime is found to have periodic attractors for certain values of the dissipation and inertia parameters.  For the aerosol regime of the two-action ABC map, an attractor merging and widening crises is identified using the bifurcation diagram and the spectrum of Lyapunov exponents. After the crisis an attractor with two parts is seen,  and trajectories hop between these parts with period 2. The bubble regime of the embedded map shows strong hyperchaotic regions  as well as  crisis induced intermittency with  characteristic times between bursts  that scale as a power law behavior as a function of the dissipation parameter. Furthermore, we observe riddled basin of attraction and unstable dimension variability in the phase space in the bubble regime. The bubble regime in one-action shows similar behavior. This study of a simple model of impurity dynamics may shed light upon the transport properties of passive scalars in three dimensional flows. We also compare our results with those seen earlier in two dimensional flows.
\end{abstract}
\maketitle

\section{Introduction}

The dynamics of inertial particles, i.e. small spherical particles immersed in a  fluid flow has  been studied extensively over the years (See \cite{Cart} and references therein). Such dynamical systems have been considered to be the simplest models for the study of transport and mixing properties of impurities in fluid flows, which are of interest in a variety of practical situations. Examples include aerosols and pollutants in the atmosphere, micro-structures suspended in fluids, bubbles in liquids, slurries in industrial mixers, droplet formation by  cloud turbulence \cite{Falkovich}, planet formation around developing stars in the universe \cite {Wilkinsion} and many others. These systems exhibit rich and complex behavior, and demonstrate phenomena that are interesting from the point of view of dynamical systems theory.

The Lagrangian dynamics of small spherical tracers in nonuniform and incompressible flows is described by the Maxey-Riley equation \cite{Maxey}, under the assumption that the fluid around the tracers is locally incompressible and of uniform density. For neutrally buoyant tracers, after various approximations \cite{Babiano}, the Maxey-Riley equation leads to a set of minimal equations known as the embedding equations \cite{Babiano,Cartwright}.  The same procedure  has been generalized for the case of non-neutrally buoyant tracers  when the particle density differs from the fluid density. The fluid dynamics is actually embedded in the particle dynamics and may be recovered under appropriate limits. The particle motion in the flow turns out to be compressible and thereby gives rise to regions of contraction and expansion without affecting the incompressible nature of the Lagrangian fluid flow. Thus, the embedding dynamics is dissipative in nature. Map analogues of the embedding equation, which preserve the features of the embedding dynamics  have also  been  constructed \cite{Motter}. In the case where the densities of the fluid and the particles do not match, their trajectories are expected to differ. This can have interesting consequences. It has been observed that in two-dimensional  chaotic flows, aerosols are pushed out from the KAM (Kolomogrov-Arnold-Moser) islands. The opposite tendency has been reported for bubbles \cite{Crisanti,Tanga, Piro}. However, neutrally buoyant particles may get detached from the fluid-parcel trajectories, and settle in the KAM islands. This behavior suggested an interesting application, a method of targeting KAM islands in Hamiltonian flows. The generalization of this method has been studied for Hamiltonian maps as well, and has been called the method of the bailout embedding \cite{Babiano,Cartwright}. It may be noted that the effects of the Basset-Boussinesq history term has been recently considered for the sedimentation and rising of inertial particles in a $2d$ convective, time-periodic, cell flow \cite{Guseva}. 

Bail-out embeddings of two dimensional flows, as well as maps have been studied earlier \cite{Cartwright, Nirmal}. The study of the bail-out embeddings of two-dimensional area-preserving maps \cite{Nirmal} indicated that the dissipation parameter of the system, as well as the density difference plays a crucial role in the dynamical behaviors of bubbles and aerosols. It has been demonstrated that the embedding map can target periodic orbits and chaotic behavior in both the aerosol and bubble regimes depending on values of the dissipation parameter $\gamma$, and the mass ratio parameter $\alpha$. Moreover, an attractor merging and widening crisis was observed in a certain parameter region for the aerosol.  Crisis induced intermittency was also found at some points in the parameter space. Here we study the bail - out embedding of a $3-d$ volume preserving map. It is expected that the presence of an extra dimension in the phase space will have important contributions to the clustering and transport properties seen in the phase space. 

In this paper we investigate the behavior of the bail-out embedding of a volume preserving $3-d$ map, the ABC map \cite{Dombre}. ABC flows \cite{Arnold} are important models in the context of magnetohydrodynamics, and can sustain the dynamo effect, i.e. a magnetic field can be generated and maintained by the motion of an electrically conducting fluid. The flows can show streamlines and periodic behavior as well as chaotic regimes accessed by a series of bifurcations \cite{Feudel}. The spatial structure of the magnetic field is influenced by the velocity field. The map analogue of the ABC flow \cite{Feingold} demonstrates similar effects. The velocity field of the ABC map has been studied both analytically and numerically in detail by Dombre \textit{et al.} \cite{Dombre}. The bail-out embedding of this volume preserving map has also been studied by Cartwright \textit{et al.} \cite{Cartwright} for neutral particles, and demonstrates the accumulation of impurities in tubular vortical structures, the detachment of particles from fluid parcels near hyperbolic invariant lines, and the formation of $3-d$ structures. Our study of the ABC map examines spherical tracers whose density is different from that of the fluid. We discuss both the two-action and one-action cases of the map, but concentrate on the two-action case here. For this case, as in the $2-d$ case, the aerosols and bubbles show distinctly different behavior. The system has a rich phase diagram wherein periodic structures, hyperchaotic regimes and chaotic behavior is seen, depending on the dissipation parameter, and buoyancy effects. We also observe unstable dimension variability (UDV) leading to crisis in this system in the aerosol regime, for certain values of the dissipation and buoyancy parameters. The system also shows a riddled basin of attraction and UDV in the bubble regime. This behavior can have interesting practical consequences.

The paper is organized in the following way. The bailout embedding method is outlined  in sec. II. The embedded version of the base $ABC$ map is constructed in section III. The aerosol and bubble regimes in the two-action case are investigated in section IV. The phase diagram of the two-action case is discussed in section V and compared with the phase diagram of the $2-d$ standard map case studied earlier.  We discuss the one-action case briefly in section VI. The conclusions are summarized in section VII.

\section{The Bailout Embedding Map}
The transport of passive point particle tracers in the fluids is usually studied using the Lagrangian framework, wherein
\begin{eqnarray}
\dot{x}=u_{x}(x,y,z,t)\nonumber \\
\dot{y}=u_{y}(x,y,z,t)\nonumber \\
\dot{z}=u_{z}(x,y,z,t)
\end{eqnarray}   
Here $\dot{x}$, $\dot{y}$ and $\dot{z}$ are the particle velocities, and $u_{x}$, $u_{y}$ and $u_{z}$ are the components of the fluid velocity field $u$. The obvious advantage of working in this framework is that the particle advection problem is now expressed as a finite dimensional dynamical system.
If the density of the particle  tracers is different from that of the fluid, then the problem needs to be tackled using the Maxey-Riley framework. In the Lagrangian description, the advection of spherical particles in an incompressible fluid is given by the Maxey-Riley equation \cite{Maxey} which has the form:

\begin{eqnarray}
\rho_{p}\frac{d\textbf{v}}{dt}=\rho_{f}\frac{d\textbf{u}}{dt}+(\rho_{p}-\rho_{f})g-\frac{9\nu\rho_{f}}{2a^{2}}(\textbf{v}-\textbf{u}-\frac{a^2}{6}\nabla^2\textbf{u})-\frac{\rho_{f}}{2}(\frac{d\textbf{v}}{dt} - \frac{D}{Dt}[\textbf{u}+\frac{a^2}{10}\nabla^2\textbf{u}]) \nonumber \\
-\frac{9\rho_{f}}{2a}\sqrt{\frac{\nu}{\pi}}\int_{0}^{t} {\frac{1}{\sqrt{(t-\xi)}}\frac{d}{d\xi}(\textbf{v}-\textbf{u}-\frac{a^2}{6}\nabla^2\textbf{u}) d\xi}
\label{Maxey-Riley}
\end{eqnarray}

Here \textbf{v} represents the particle velocity, \textbf{u} the fluid velocity, $\rho_{p}$ the density of the particle, $\rho_{f}$  the density of the fluid, and $\nu$, \textbf{a} and \textbf{g} 
represent the kinematic viscosity of the fluid, the radius of the particle and the acceleration due to gravity respectively. The first term on the right of Eq. (\ref{Maxey-Riley}) represents the force exerted by the undisturbed flow on the particle, the second term represents the buoyancy, the third term represents the Stokes drag, the fourth term represents the added mass, and the last term is the Basset-Boussinesq history force term. The derivative $\frac{D \textbf{u}}{dt}$ where $\frac{D \textbf{u}}{Dt} = \frac {\partial \textbf{u}}{\partial t}+(\textbf{u}. \nabla)\textbf{u}$ is taken along the path of the fluid element, and the derivative  $\frac{d\textbf{u}}{dt}=\frac{\partial\textbf{u}}{\partial t} + (\textbf{v}.\nabla)\textbf{u}$ is taken along the trajectory of the particle.
The Maxey-Riley equation is derived under the assumption that the particle radius, and the Reynolds number are small, and so are  the velocity gradients around the particle. It is also assumed that the initial velocities of the particle and fluid are the same. A full review of the problem is found in \cite{Michaelidis}.

Under the low Reynolds number approximation, with negligible buoyancy effects,  and retaining only the Bernoulli, Stokes drag and Taylor added terms, we arrive at the  following simplified equation of motion for the motion of a spherical particle immersed in the fluid \cite{Babiano}:
\begin{equation}
\rho_{p}\frac{d\textbf{v}}{dt} = \rho_{f}\frac{d\textbf{u}}{dt}-\frac{9\nu\rho_{f}}{2a^{2}}(\textbf{v}-\textbf{u})-\frac{\rho_{f}}{2}(\frac{d\textbf{v}}{dt}-\frac{d\textbf{u}}{dt})
\end{equation}

This equation has been derived under approximation that $\frac{Du}{Dt}= \frac{du}{dt}$. Expressing the equation in a non-dimensional form by proper rescaling of length, time and velocity by scale factors $L$,$T = L/U$, and $U$, we obtain
\begin{equation}
\frac{d\textbf{v}}{dt} - \alpha\frac{d\textbf{u}}{dt} = -\frac{2}{3}(\frac{9\alpha}{2a^{2}Re})(\textbf{v}-\textbf{u})
\end{equation}
Here, the parameter $\alpha$ gives the mass ratio, $\alpha = 3\rho_{f}/(\rho_{f}+2\rho_{p})$. 
Hence, values of $\alpha < 1 $  correspond  to the  aerosol case and the values $\alpha>1$, $\alpha=1$ corresponds to the bubble case, and the neutrally buoyant cases.
Defining the particle Stoke's number $St=\frac{2}{9}a^2 Re$ and defining the dissipation parameter to be $\gamma=\frac{2 \alpha}{3 St}$, the  equation takes the form\cite{Babiano}, 
\begin{equation}
\frac{d\textbf{v}}{dt} - \alpha\frac{d\textbf{u}}{dt} = -\gamma(\textbf{v}-\textbf{u})
\label{bail0}
\end{equation}
Let the base flow be represented by an area or volume preserving map, with the map evolution equation $x_{n+1}=T(x_{n})$. 
A general bailout embedding for this map which represents the motion of the particle and encapsulates the essential features of  Eq. (\ref{bail0}) may be constructed \cite{Cartwright}:
\begin{equation}
x_{n+2}-T(x_{n+1})=K(x_{n})(x_{n+1}-T(x_{n}))
\end{equation}
In the case of inertial particles, the bailout embedding function  $K(x)$ may be chosen such that the embedding map takes the form \cite{Motter}:
\begin{equation}
x_{n+2}-T(x_{n+1})=e^{-\gamma}(\alpha x_{n+1}-T(x_{n}))
\end{equation}
This can be expressed as
\begin{eqnarray}
x_{n+1} = T(x_{n})+\delta_{n} \nonumber\\
\delta_{n+1}=e^{-\gamma}[\alpha x_{n+1}-T(x_{n})]
\end{eqnarray}
This is the bailout embedding map. The dissipation parameter $\gamma$ is a measure of contraction or expansion in the phase space of the particle's dynamics. The particle is said to have bailed out of the fluid trajectory when $\delta \neq 0$, where the new variable $\delta$ is  defined to be the detachment of the particle from the local fluid parcel. The fluid dynamics is recovered under the limits $\delta \rightarrow 0 $, $\alpha = 1$ and $\gamma \rightarrow \infty$. In this sense, the fluid dynamics is said to be embedded in the particle's equation. This map is dissipative with a phase space contraction rate which is $e^{-\gamma}$.  The configuration space contraction rate is proportional to $e^{-\gamma}(\alpha-1)$ for $e^{-\gamma}(\alpha-1) << 1$ \cite{Motter}. We will use this version of the bail-out embedding map in all subsequent analysis.

\section{The Embedded Arnold-Beltrami-Childress (ABC) Map} 
We consider the Arnold-Beltrami-Childress (ABC) Map \cite{Dombre,Feingold} as the base map for the present study. The ABC map exhibits almost all the basic features of interest in the evolution of a typical three dimensional, time periodic,  volume preserving flow, the ABC flow.
This class of simple  non-turbulent flows, first introduced by Arnold is known to possess KAM-like surfaces and  generate chaotic streamlines. In  the  chaotic regions,  the flow has the Beltrami property, i.e. the vorticity is parallel to the fluid velocity. The passive scalars immersed in such streamlines display chaotic advection, which considerably enhances the mixing and transport properties of passive scalars, as shown in the two-dimensional case of the blinking vortex model introduced by Aref  \cite{Aref}. Moreover, Childress \cite{Childress} demonstrated that the presence of chaotic streamlines in the flow is responsible for the growth of the magnetic field in the model of the Kinematic Dynamo Effect.
We use the following version of the ABC Map \cite{Dombre} for our purpose:
\begin{empheq}[right=\empheqrbrace \mod 2\pi]{align}
x_{n+1}=x_{n}+A\sin (z_{n}) + C\cos(y_{n})  \nonumber\\
y_{n+1}=y_{n}+B\sin(x_{n+1}) +A\cos(z_{n})  \nonumber\\
z_{n+1}=z_{n}+C\sin(y_{n+1}) + B\cos(x_{n+1})
\end{empheq}

\begin{figure*}
\begin{tabular}{cc}
 \includegraphics[height=7cm,width=8.5cm]{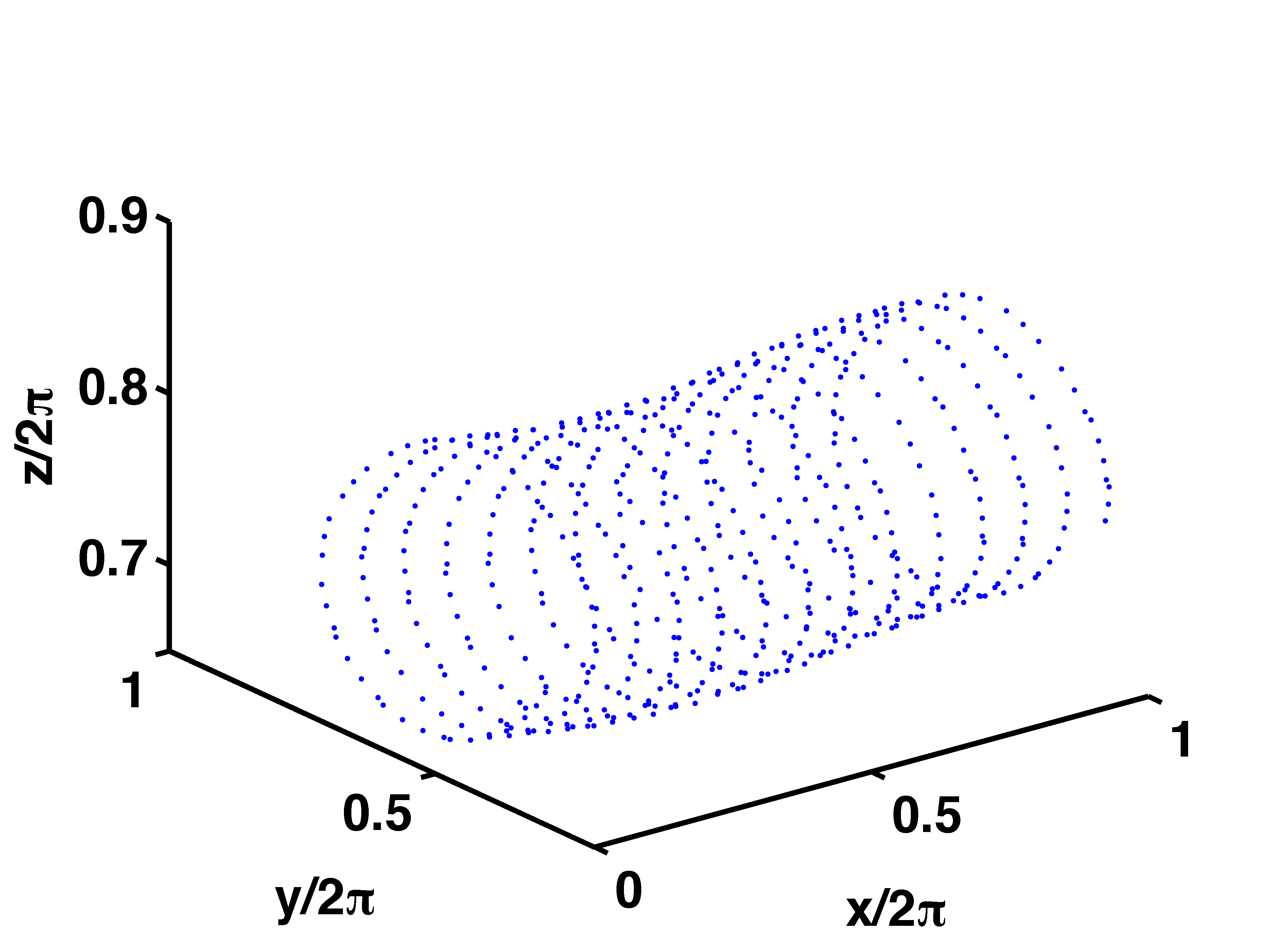}&
\includegraphics[height=7cm,width=8.5cm]{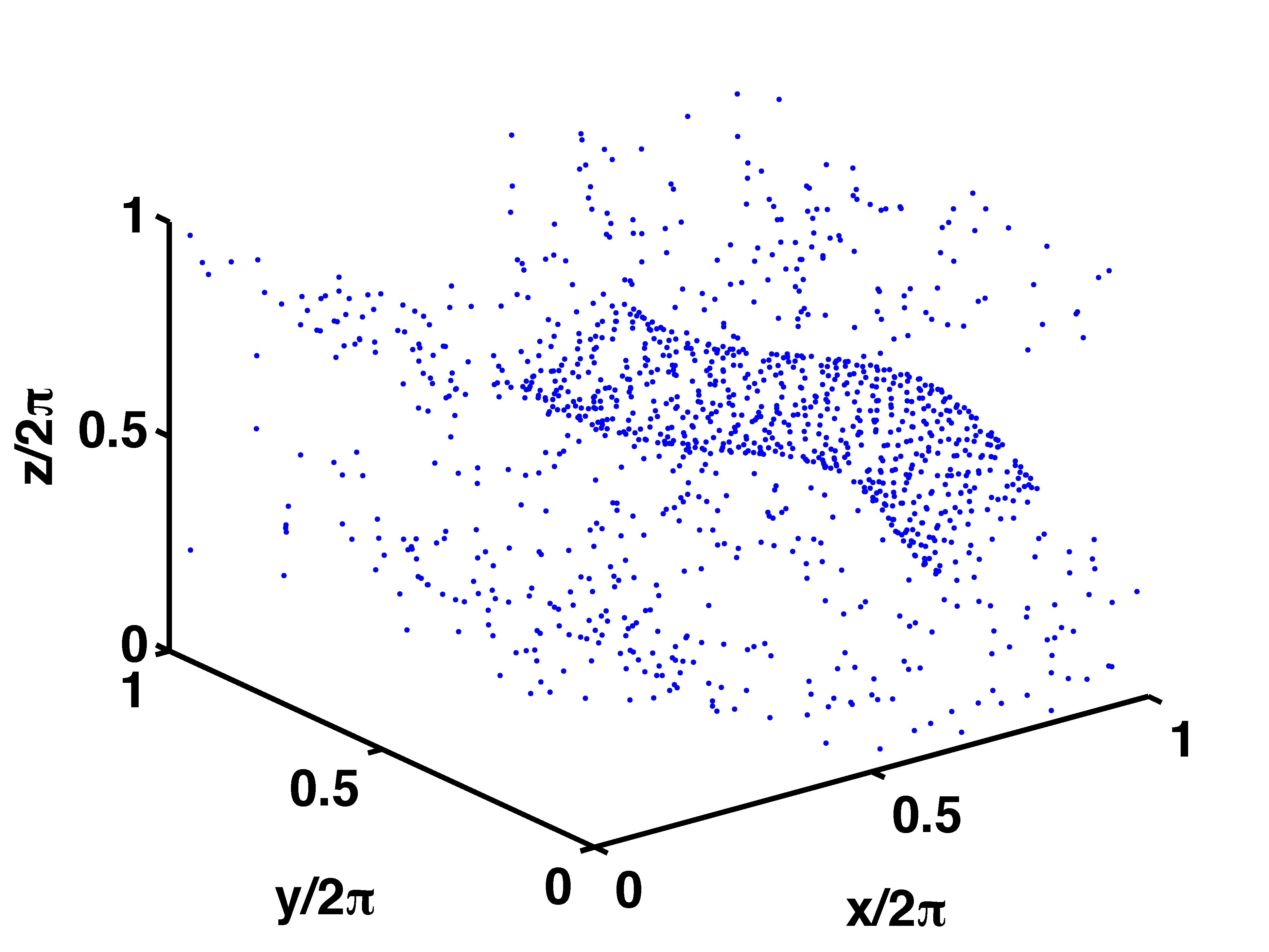}\\
(a) & (b)
\end{tabular}{}
\caption{\label{fig:ABC} \footnotesize The x-y-z phase space of the ABC map in (a) the one-action case  for the parameter values (A,B,C)=(1.5,0.08,0.16), and (b) the  two-action case for the parameter values (A,B,C)=(2.0,1.5,0.08).} 
\end{figure*}

The map is implemented modulo $2\pi$ with real parameters $(A,B,C)$. There are two quasi-integrable cases of the ABC map \cite{Integrability}. However, it is the presence of the  chaotic streamlines seen in the non-integrable case that makes it an interesting prototype to be used for fluid dynamical studies in three-dimensions. The map is referred to as the one-action map if one of the parameters exceeds 1; and as the two-action version if two of parameters $(A,B,C)$ are larger than 1. The one-action ABC map shows KAM-like invariant surfaces, parts of  which break down under small perturbations but the trajectories  remain bounded within the invariant surfaces that are intact (see Fig.~\ref{fig:ABC}(a)). In the two-action ABC map, however, unbounded diffusive motion through the invariant surfaces is found (see Fig.~\ref{fig:ABC}(b)). Consequently, the trajectories fills up all the accessible phase space. This phenomenon is known as  resonance-induced diffusion. Figs.~\ref{fig:ABC} (a) and (b) show the $3-d$ phase space of the ABC map for the one action and the two action cases for the parameter values $ A=1.5, B=0.08, C=0.16$ (one action) and for the parameter values $A=2,B = 1.5, C = 0.08$ (two action), respectively. In Fig.~\ref{fig:ABC}(a) at the parameter values $(A,B,C)$=$(1.5,0.08,0.16)$, we see a tube like KAM surface typical of those seen for one-action maps \cite{Dombre} whereas  Fig.~\ref{fig:ABC}(b) at the parameter values $(A,B,C)$=$(2.0,1.5,0.16)$, clearly shows chaotic regions and KAM tubes separated by invariant surfaces that prohibit any transport. The phase diagram of the embedded system in the $ \alpha$ - $\gamma$ space for the two-action case shows many interesting regimes which will be discussed in a subsequent section.

The bailout embedded version of the ABC map is given by the following 6-dimensional map:

\begin{empheq}[right=\empheqrbrace \mod 2\pi]{align}
x_{n+1}=x_{n}+A\sin(z_{n}) + C\cos(y_{n}) +\delta_{n}^{x} \nonumber\\
y_{n+1}=y_{n}+B\sin(x_{n+1}) + A\cos( z_{n})+\delta_{n}^{y} \nonumber\\
z_{n+1}=z_{n}+C\sin(y_{n+1}) + B\cos(x_{n+1}) +\delta_{n}^{z} \nonumber\\
\delta_{n+1}^{x} = e^{-\gamma}[\alpha x_{n+1}-(x_{n+1} - \delta_{n}^{x})] \nonumber\\
\delta_{n+1}^{y} = e^{-\gamma}[\alpha y_{n+1}-(y_{n+1} - \delta_{n}^{y})] \nonumber\\
\delta_{n+1}^{z} = e^{-\gamma}[\alpha z_{n+1}-(z_{n+1} - \delta_{n}^{z})] 
\end{empheq}

This is a dissipative and invertible map implemented modulo $2\pi$. The case of neutrally buoyant particles in ABC maps and flows has been studied earlier \cite{Cartwright}. Here, it was seen that in the one-action case, particles were expelled from the chaotic regions to finally settle into the tubular KAM structures. In the two-action case, the neutral particles and the fluid parcels followed exponentially convergent trajectories and small fluctuations could be induced by the presence of noise. Here, we consider the case where the particles and the fluid have different densities. As mentioned earlier, the parameters $\alpha$ and $\gamma$ are the inertia and dissipation parameters respectively. We consider the aerosol regime [$\alpha<1$] and bubble regime [$(1<\alpha<3)$] in detail in the following sections.  We first discuss the two-action ABC map case. The one action case is addressed in a later section.

\section{The two-action case}
\begin{figure*}
\begin{tabular}{cc}
 \includegraphics[height=7cm,width=8.5cm]{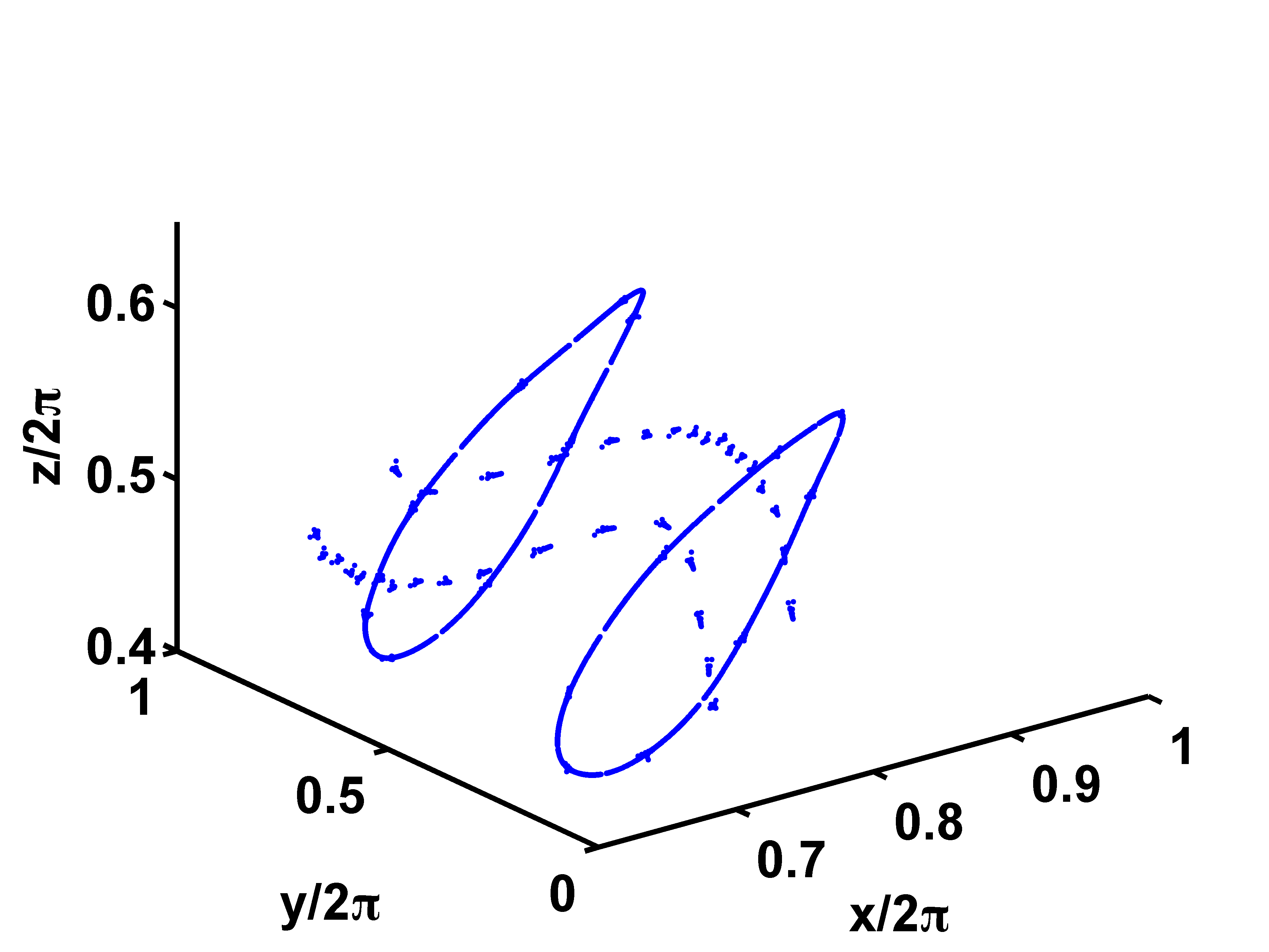}&
\includegraphics[height=7cm,width=8.5cm]{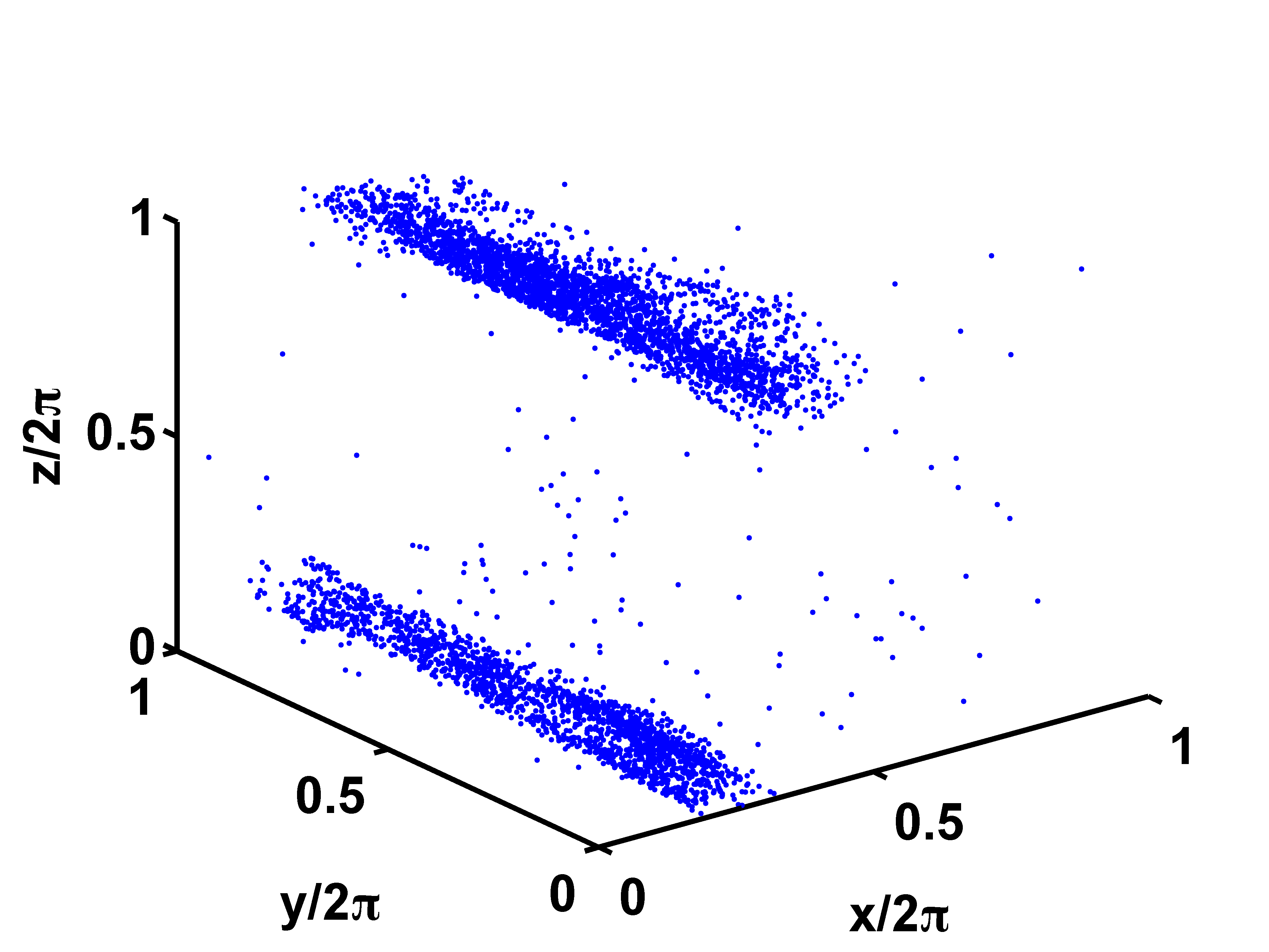}\\
(a) & (b)
\end{tabular}{}
\caption{\label{fig:embed_two_ABC} \footnotesize (Color online) The x-y-z phase space of the embedded two-action ABC map in (a) the aerosol regime for the parameter values (A,B,C)=(2.0,1.5,0.08) and $ (\alpha,\gamma) = (0.7,3.6)$, and (b) the bubble regime for the parameter values (A,B,C)=(2.0,1.5,0.08) and $ (\alpha,\gamma) = (2.0, 2.2).$}
\end{figure*}

The embedded two action $ABC$ map contains five parameters $A,B,C, \alpha$,
and $\gamma$, and shows many different kinds of behavior depending on the parameter values.
Here, we examine the behavior of the embedded two-action map at two parameter sets where interesting structures are seen. Fig.~\ref{fig:embed_two_ABC}(a) shows the behavior of the particles in the fluid for the parameter values $A=2, B=1.5, C=0.08$, and $\alpha=0.5$, $\gamma=3.6$. This is the aerosol regime, and the aerosols evolve to form a pair of ring like structures. The trajectory points which do not lie on the rings constitute the transient to the asymptote seen in the Fig.~\ref{fig:embed_two_ABC}(a). This is a regime with strong dissipation. On the other hand, in the bubble regime, with $\alpha=2.0$, $\gamma=2.2$, the bubbles end up on two raft like structures. There is also a random distribution of points in between.  We note that many other structures are possible at other parameter values. This will be discussed in detail when we discuss the phase diagram. 

\subsection{The aerosol regime : Interior crisis and unstable dimension variability}
\begin{figure*}[hbtp]
\centering
\includegraphics[height=7cm,width=8.5cm]{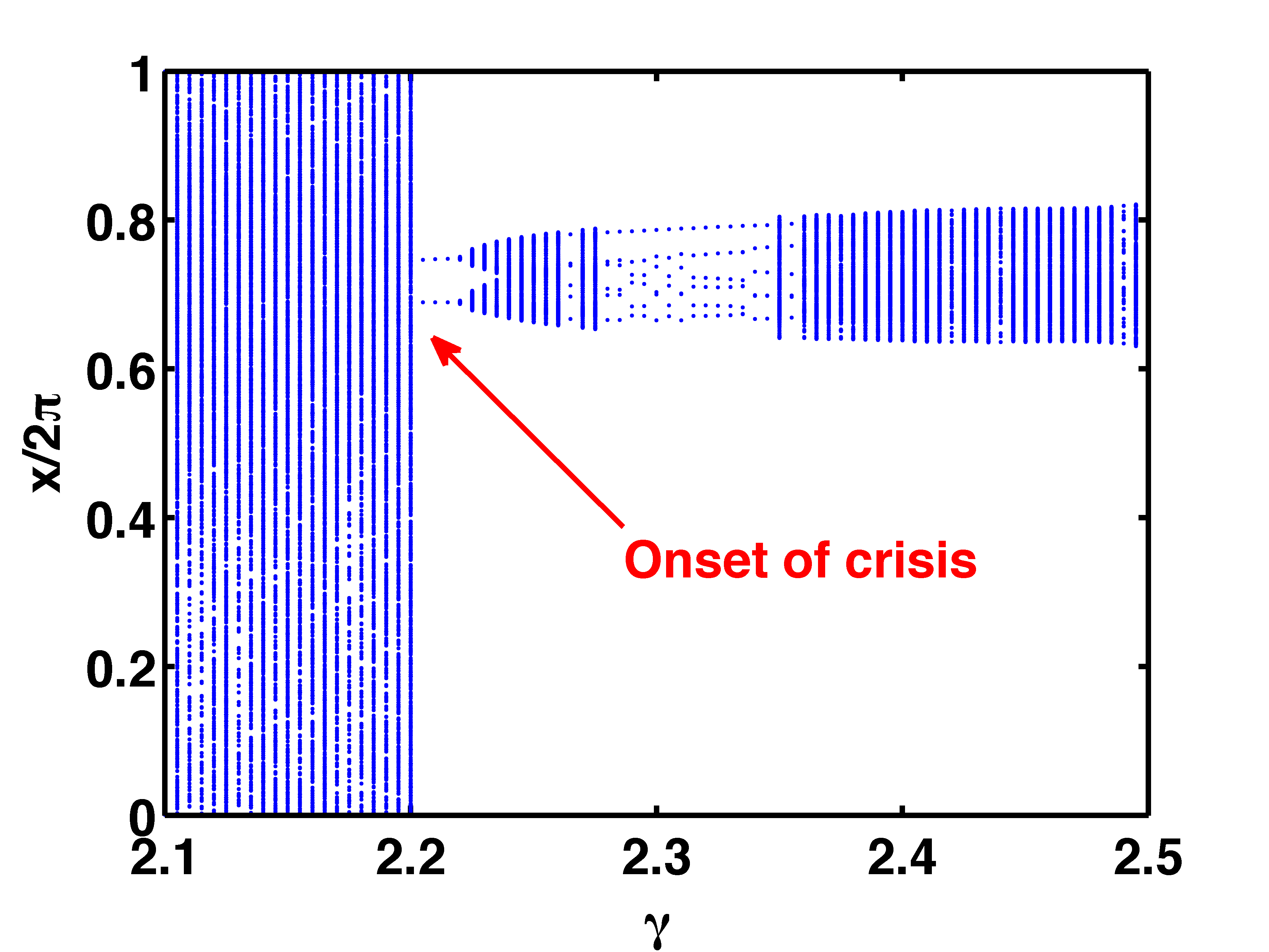}
\caption{\footnotesize\label{fig:Bifurcation aerosol} (Color online) The bifurcation diagram for the embedded two-action ABC map in the aerosol regime for the parameter values (A,B,C) =(2.0,1.5,0.16) and $\alpha = 0.7$.}
\end{figure*}

\begin{figure*}
\begin{center}
\begin{tabular}{cc}
 \includegraphics[height=7cm,width=8.5cm]{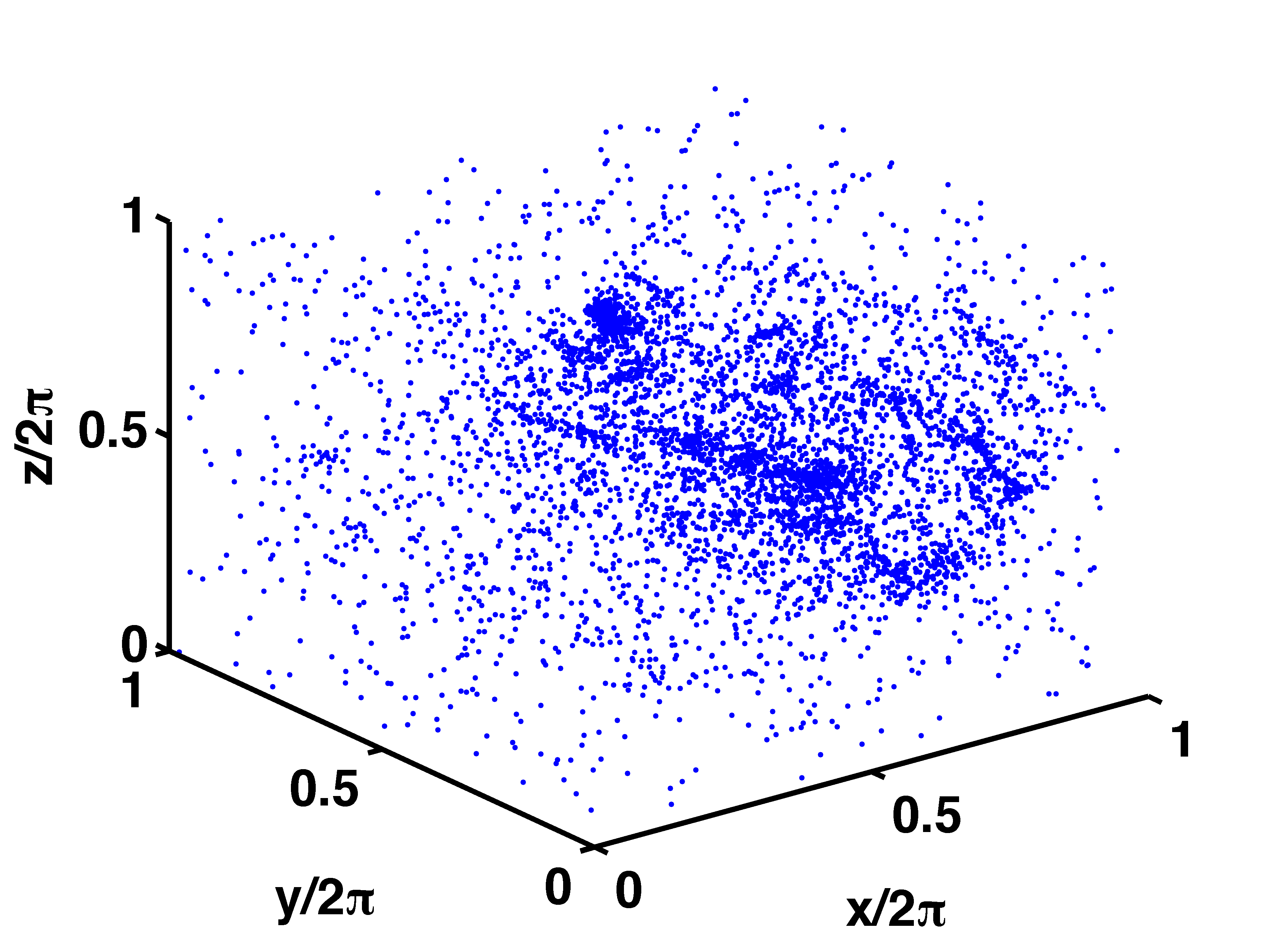}&
\includegraphics[height=7cm,width=8.5cm]{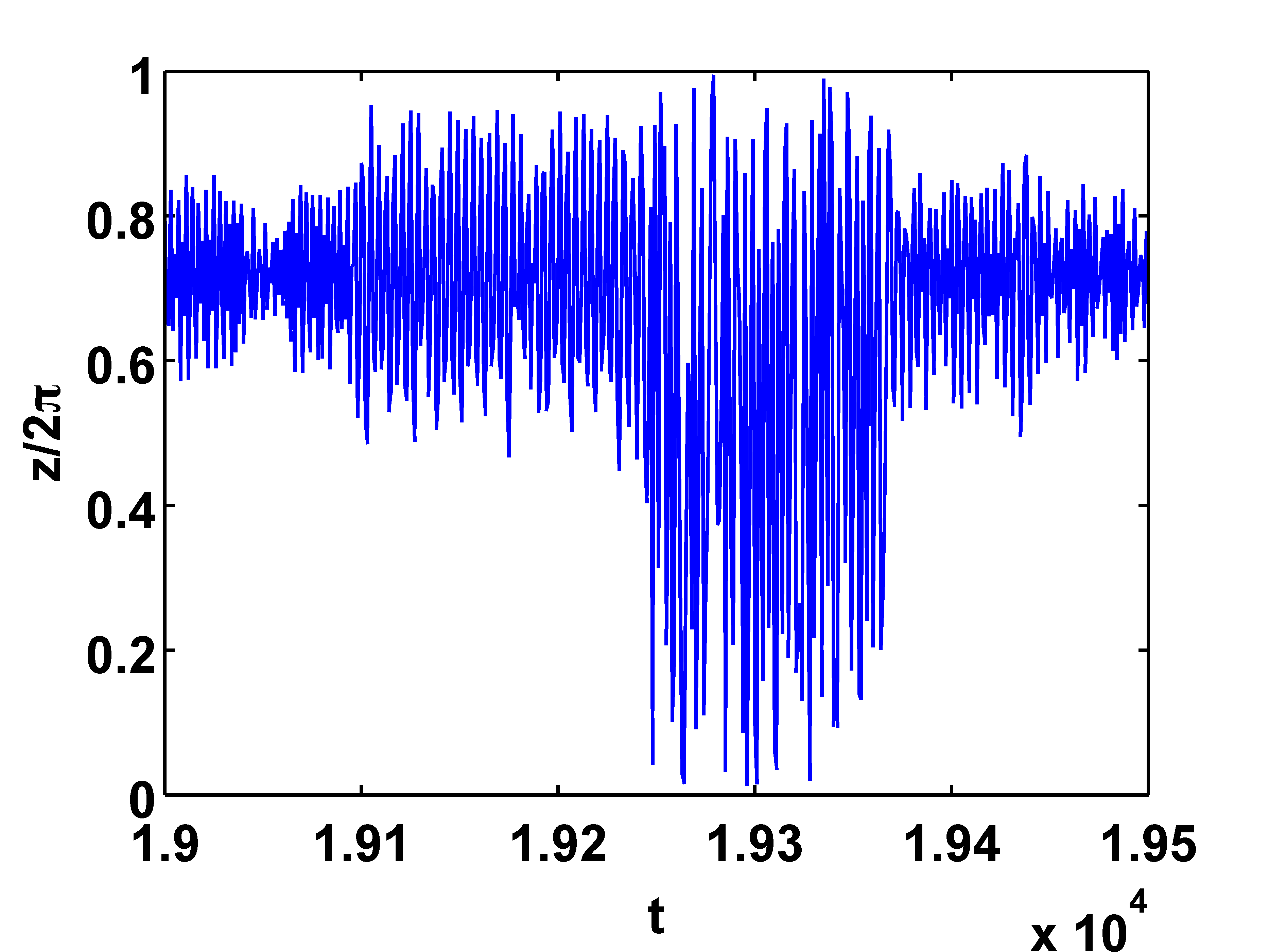}\\
(a) &(b) \\
\end{tabular}{}
\caption{\footnotesize\label{fig:Pre_crisis_Aero} (Color online) The pre crisis scenario for the embedded two-action ABC map in the aerosol regime for the parameter values (A,B,C)=(2.0,1.5,0.16) and $ (\alpha,\gamma) = (0.7, 2.2)$, (a) the attractor in x-y-z phase space  (500 transients discarded) and (b) the time series.}
\end{center}
\end{figure*}

\begin{figure*}
\begin{center}
\begin{tabular}{cc}
\includegraphics[height=7cm,width=8.5cm]{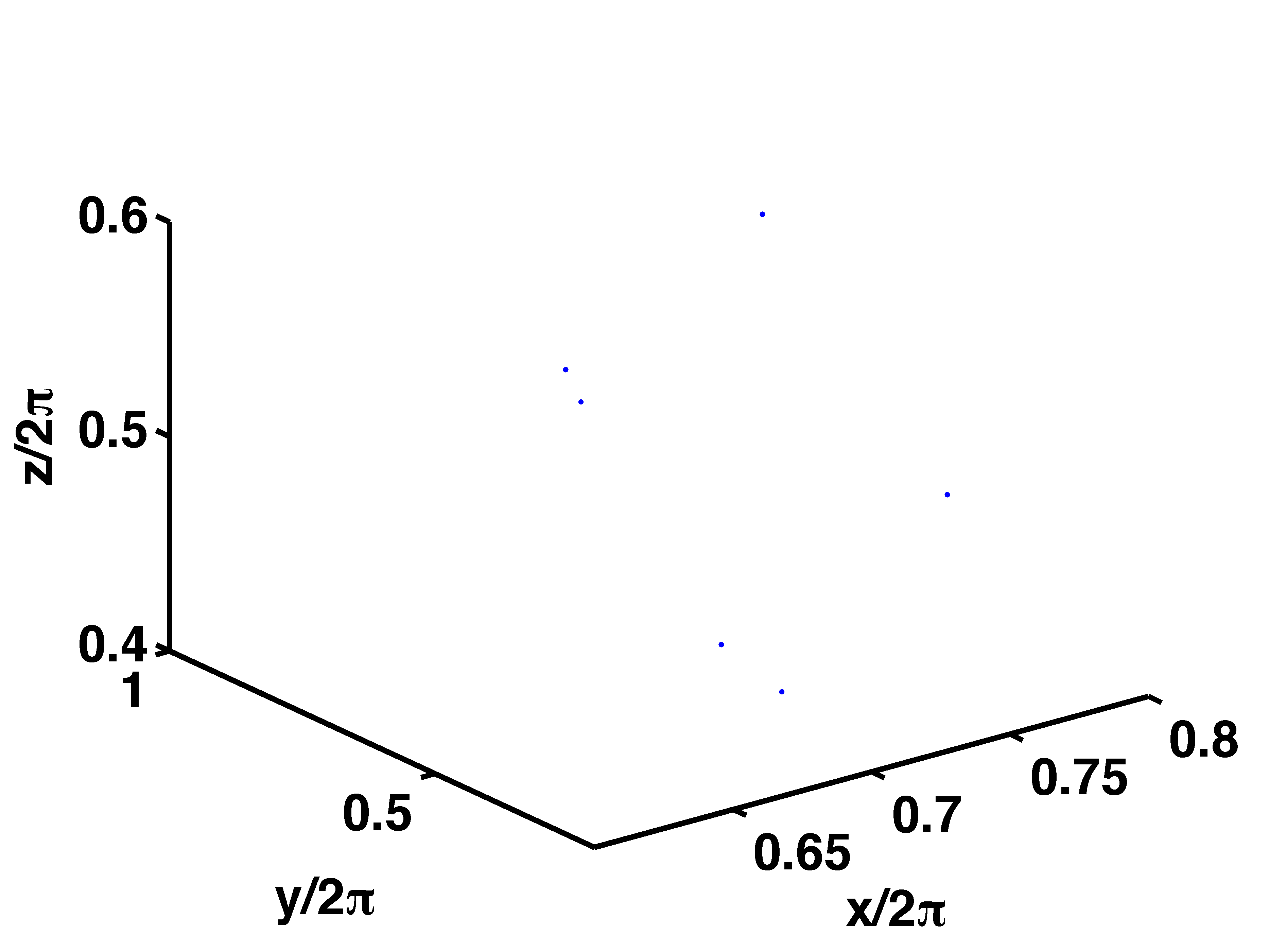}&
\includegraphics[height=7cm,width=8.5cm]{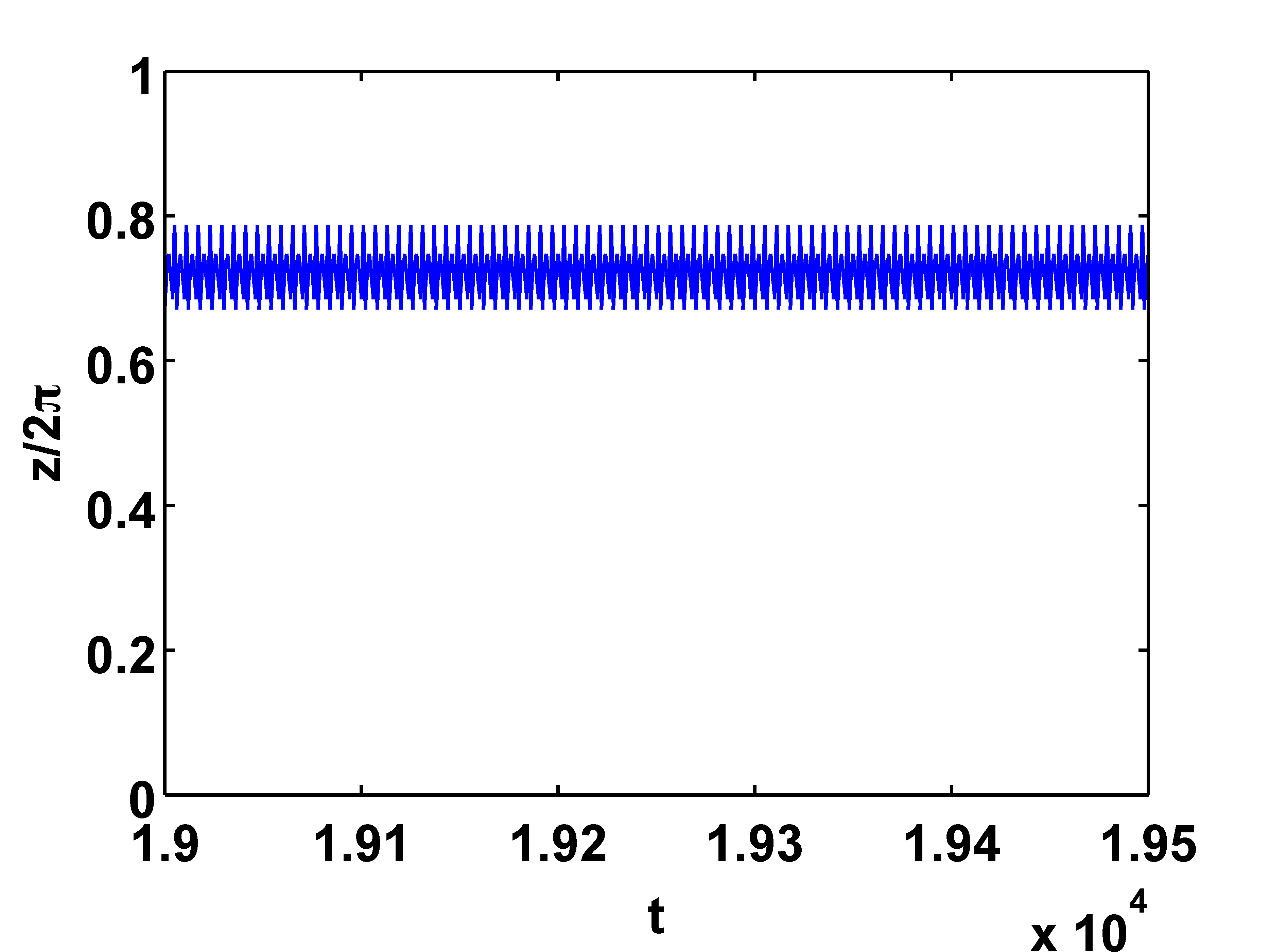}\\
(a) &(b) \\
\end{tabular}{}
\caption{\footnotesize\label{fig:Post_crisis_Aero} (Color online) The post crisis scenario for the embedded ABC map in the aerosol regime for the parameter values (A,B,C)=(2.0,1.5,0.16) and $ (\alpha,\gamma) = (0.7, 2.3)$, (a) the attractor in x-y-z phase space  (500 transients discarded) and (b) the time series}
\end{center}
\end{figure*}

\begin{figure*}
\begin{center}
\begin{tabular}{cc}
 \includegraphics[height=7cm,width=8.5cm]{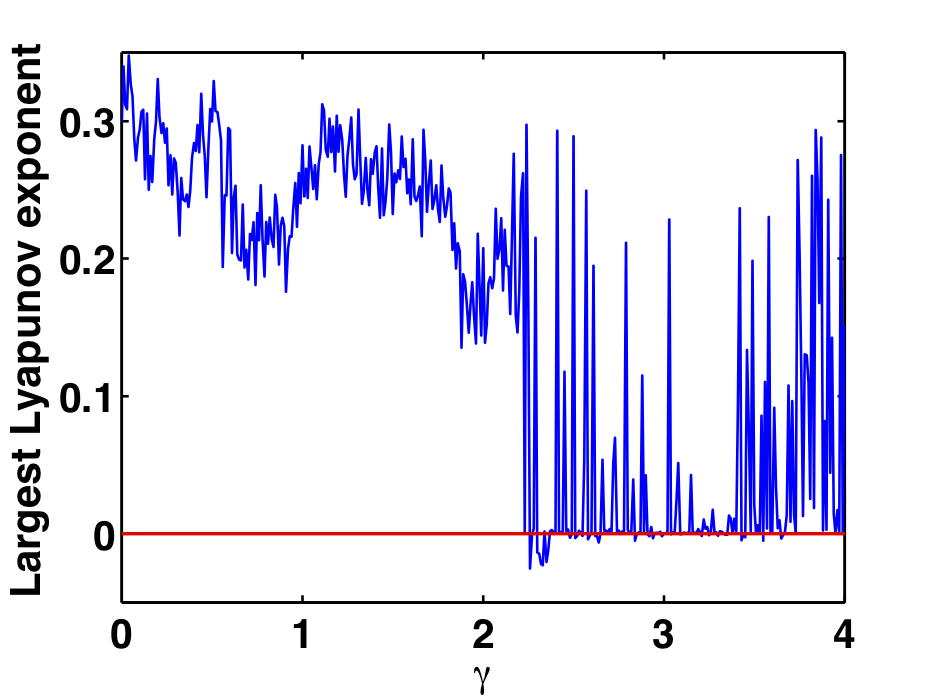}&
\includegraphics[height=7cm,width=8.5cm]{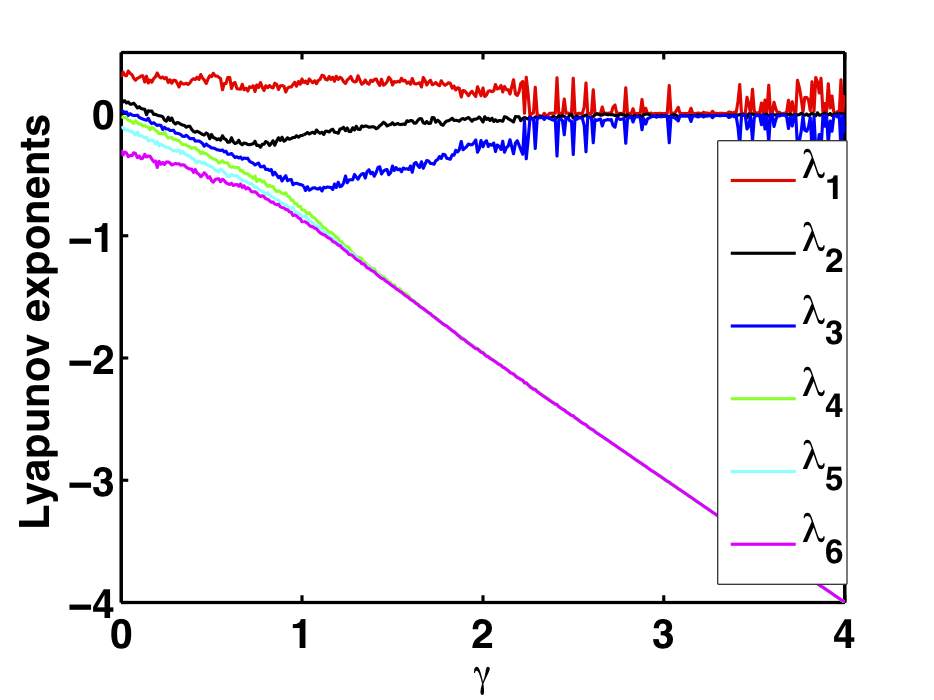}\\
(a) &(b) \\
\end{tabular}{}
\caption{\footnotesize\label{fig:Lyapunov_aero} (Color online) The Lyapunov exponents for the embedded two-action ABC map in the aerosol regime for the parameter values  (A,B,C)=(2.0,1.5,0.16) and $\alpha = 0.7$, (a) The largest Lyapnunov exponent and (b) The full spectrum ($\lambda_{1} > \lambda_{2} >\lambda_{3} >\lambda_{4} >\lambda_{5} >\lambda_{6}$).}
\end{center}
\end{figure*}

\begin{figure*}[hbtp]
\centering
\includegraphics[height=7cm,width=8.5cm]{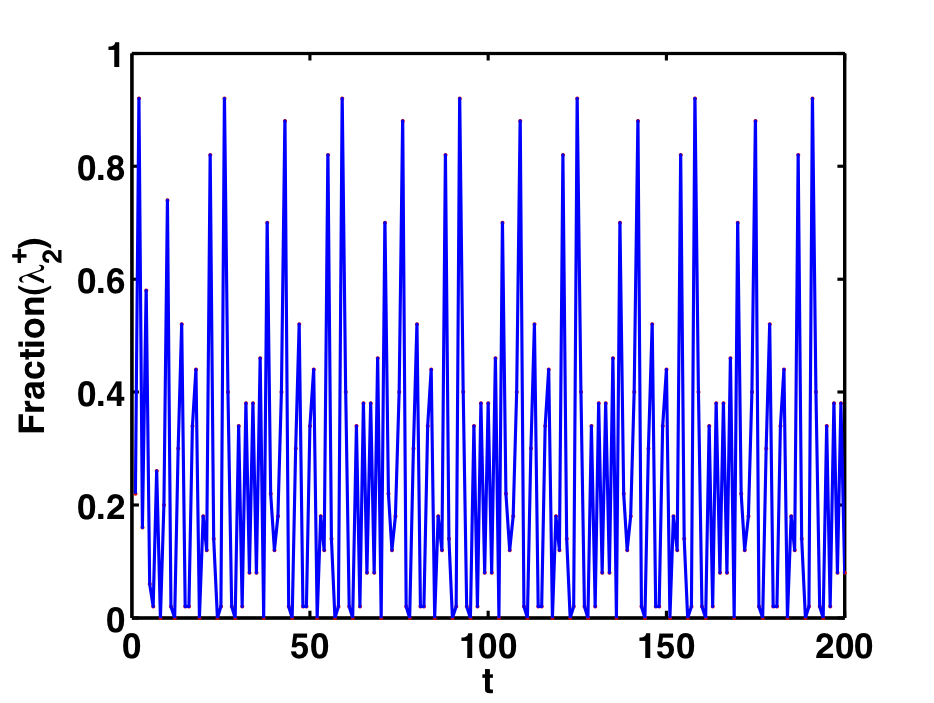}
\caption{\footnotesize\label{fig:UDV_aero} (Color online)Plot of the fraction of positive time-50 Lyapunov exponents versus time, in the aerosol regime for the parameter values (A,B,C) =(2.0,1.5,0.16) and $(\alpha,\gamma) = (0.7,2.2)$. UDV is clearly seen.}
\end{figure*}

\begin{figure*}[hbtp]
\centering
\includegraphics[height=8.5cm,width=9.5cm]{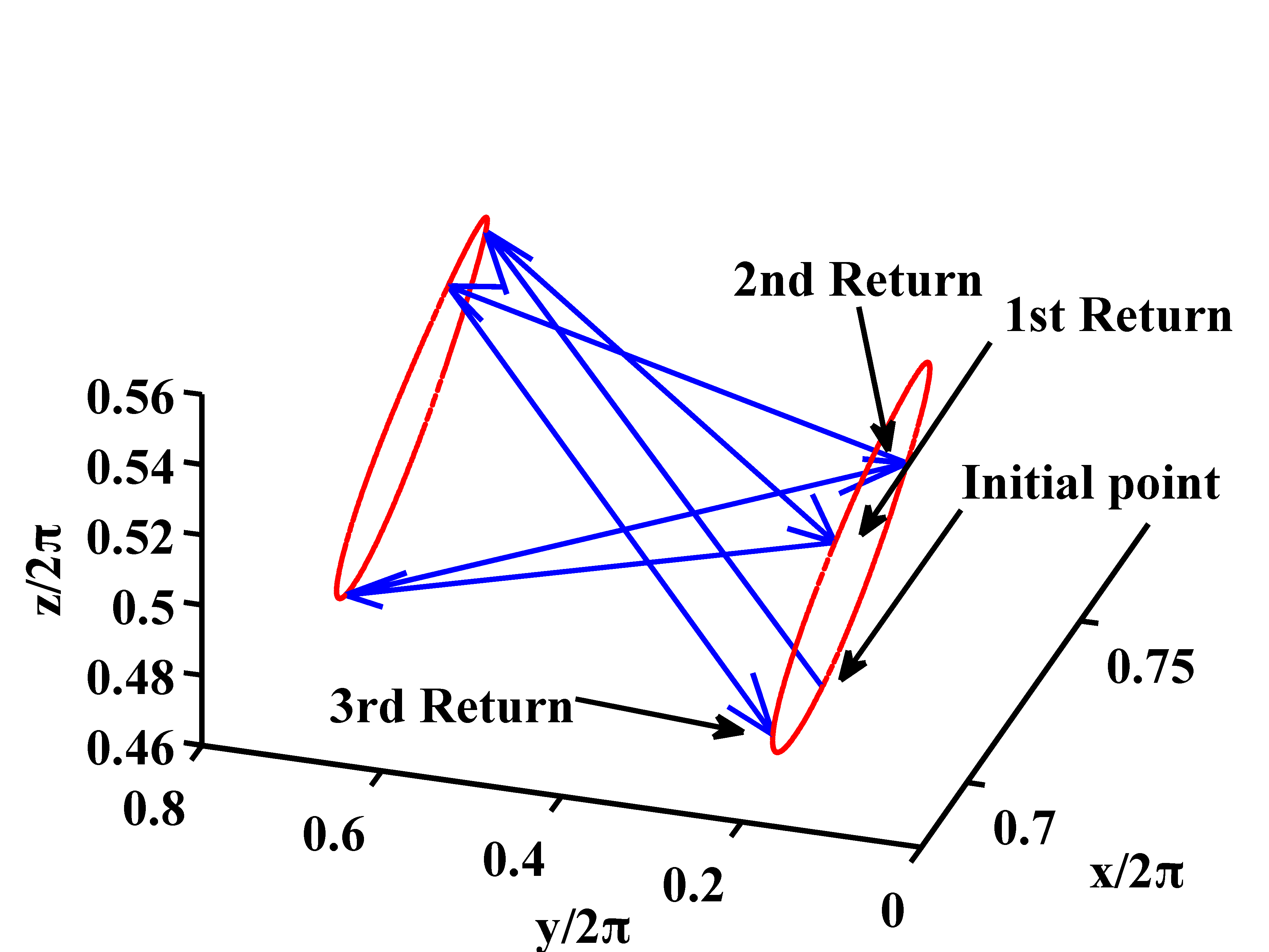}
\caption{\footnotesize\label{fig:Hopping}(Color online) The hopping in the aerosol regime for the parameter values (A,B,C) = (2.0,1.3,0.16) and $(\alpha,\gamma = 0.7,2.89)$.}
\end{figure*}

The aerosol regime of the embedding map shows rich structure. The major features of this regime can be identified by examining the bifurcation diagram (Fig.~\ref{fig:Bifurcation aerosol}). In this bifurcation diagram,  we plot the entire set of values accessed by the variable $\frac{x}{2\pi}$ at a given value of the dissipation parameter $\gamma$ for the entire range  where   $\gamma$  takes values  $2.1\leq\gamma\leq2.5$ \cite{Bifur}. Here, the phenomenon of  crisis is clearly visible at $\gamma = 2.235$ where the attractor, which initially covers the full available range of $x$, suddenly shrinks to a set of two points. In fact, there are several windows of crisis (at $\gamma=2.27$,$\gamma=2.35$) as the parameter $\gamma$ is increased to higher values. This crisis can  be identified  as an attractor merging and widening or interior crises. The pre-crisis attractor can be seen in Fig.~\ref{fig:Pre_crisis_Aero} and the post-crisis attractor can be seen in Fig.~\ref{fig:Post_crisis_Aero}. Fig.~ \ref{fig:Lyapunov_aero} shows the largest Lyapunov exponent (LE) of the system, plotted for the range $\gamma=0.0$ to $\gamma=6.0$. We note that unstable dimension variability, the consequence of a strong form of nonhyperbolicity,  is seen in the neighborhood of the crisis. The signature of the UDV is seen in the fact that the Lyapunov exponent clearly fluctuates around zero in the range $\gamma=1.9$ to $\gamma=3.0$. 
In addition to the asymptotic Lyapunov exponent, we calculate the finite time Lyapunov exponents of the system, as they sample the local stretching and contracting rates.
The fraction of positive finite time Lyapunov exponents i.e. Lyapunov exponents calculated  fluctuates as a function of time,  confirming that the number of stretching and contracting directions are changing as a function of time (i.e. the trajectory samples different numbers of stretching and contracting directions at different points in the attractor), indicating the existence of unstable dimension variability in this regime [Fig.~\ref{fig:UDV_aero}]. A similar form of UDV is also seen in the bubble regime, where we will discuss it in further detail.  Fig.~\ref{fig:Lyapunov_aero}(b) plots the full Lyapunov spectrum, i.e. all the six Lyapunov exponents of the embedded map in the $\gamma=0.0$ to $\gamma=3.0$ regime. Here, it is seen that the first and second largest Lyapunov fluctuate around zero in the neighborhood of $\gamma=0.5$, after which they separate, with the largest taking positive values, and the second taking negative values till about $\gamma=1.3$ where they start fluctuating around zero, together with the third largest LE which comes up from below and fluctuates around zero in the same regime. The next three Lyapunov exponents remain in the negative regime, indicating the existence of three stable directions in the six dimensional space. Similar behavior is observed in the bubble regime.  Post-crisis, interestingly, at the parameter values (A,B,C) = (2.0,1.3,0.16) and $(\alpha,\gamma = 0.7, 2.82)$, an attractor with two symmetric loops appears in the phase space and the asymptotic trajectory continuously hops between the loops with period 2 (Fig.~\ref{fig:Hopping}).  The size and location of the loops changes with initial conditions.  This phenomenon occurs in one of the windows of crises.

\begin{figure*}
\begin{center}
\begin{tabular}{cc}
 \includegraphics[height=7cm,width=8.5cm]{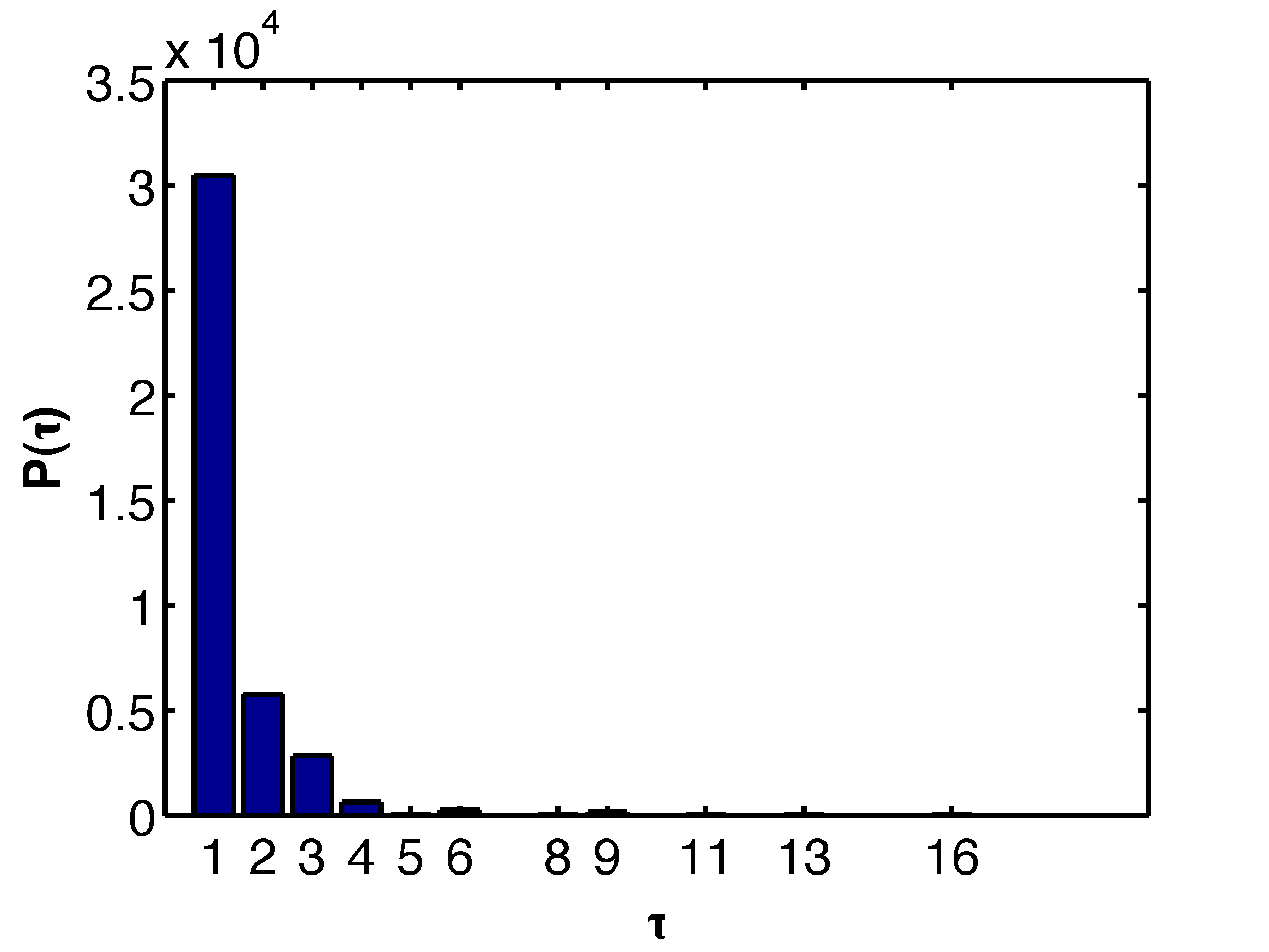}&
\includegraphics[height=7cm,width=8.5cm]{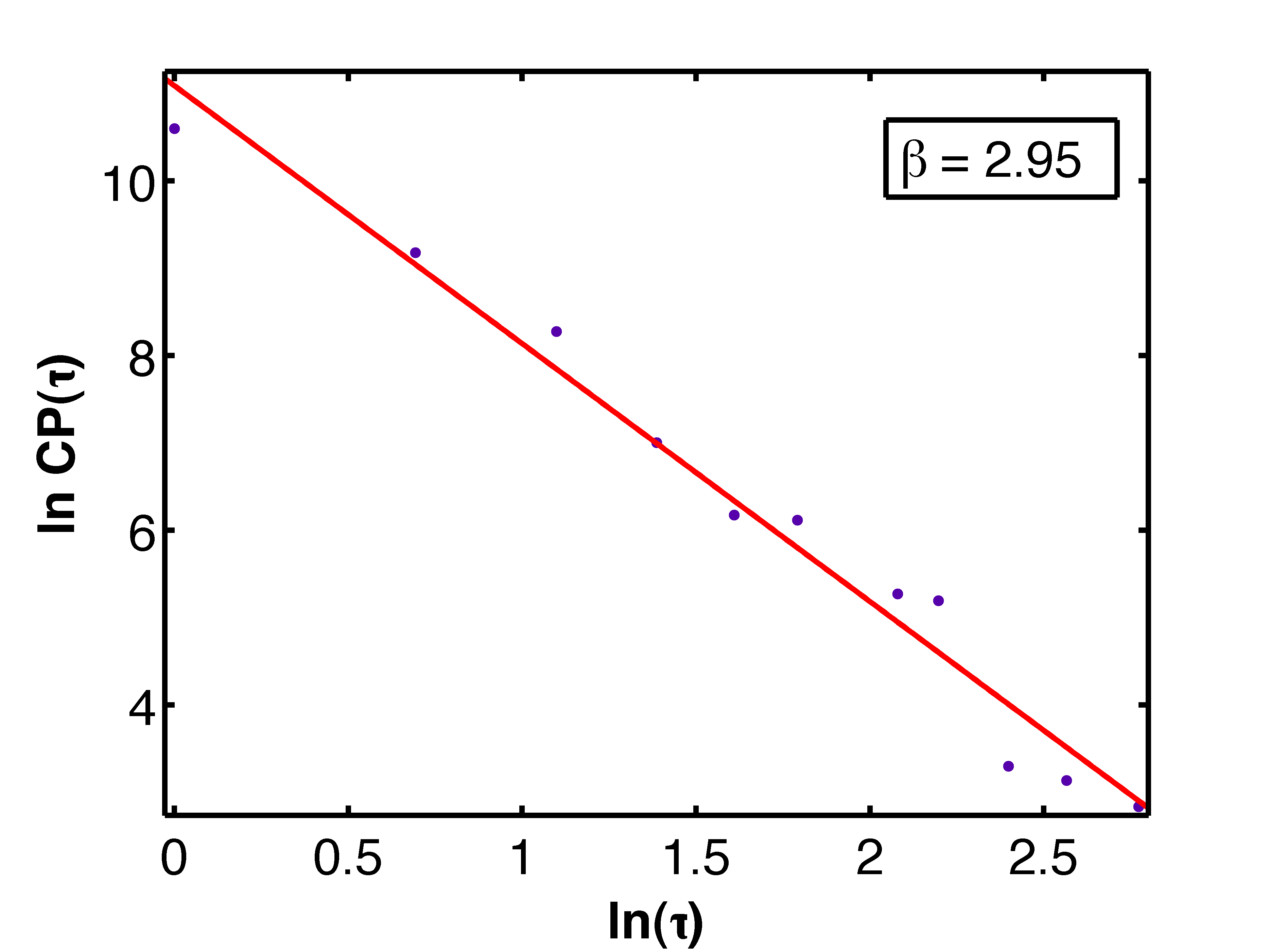}\\
(a) &(b) \\
\end{tabular}{}
\caption{\footnotesize\label{fig:Hop_stat} (Color online) Statistics of the transients before the hopping in the aerosol regime for the parameter values (A,B,C)=(2.0,1.3,0.16) and $ (\alpha,\gamma) = (0.7,2.89)$, (a) the probability distribution and, (b) the corresponding log-log plot of the reverse cumulative distribution\cite{RCD}}
\end{center}
\end{figure*}

However, the transient before the asymptote is reached shows interesting behavior. The transient tends to stick around the two rings before finally locking onto the two rings and the period two orbit. In this process, it spends some time in the neighborhood of the rings before hopping away to other region of the phase space. The distribution of the time spent  by the transients in the region of the rings is plotted in Fig.~\ref{fig:Hop_stat}(a), and the corresponding log-log plot  of the reverse cumulative distribution \cite{RCD} is shown in Fig.~\ref{fig:Hop_stat}(b). A power-law can be seen in the reverse cumulative distribution over a short regime ranging from $0.0 $ to $2.6 $ with the exponent $\beta=2.93$. The plots shown in Fig.~\ref{fig:Hop_stat} are for the transients for 500 random initial conditions and the maximum time scale here is $\tau = 16$ after which the ring attractors are reached.  A similar power law distribution has been seen in the time spent by a particle in the vicinity of a vortex in the case of Rayleigh-Ben\'{a}rd convection \cite{RBC}.

\subsection{The bubble regime : Unstable dimension variability and a riddled basin of attraction}
\begin{figure*}[hbtp]
\centering
\includegraphics[height=7cm,width=8.5cm]{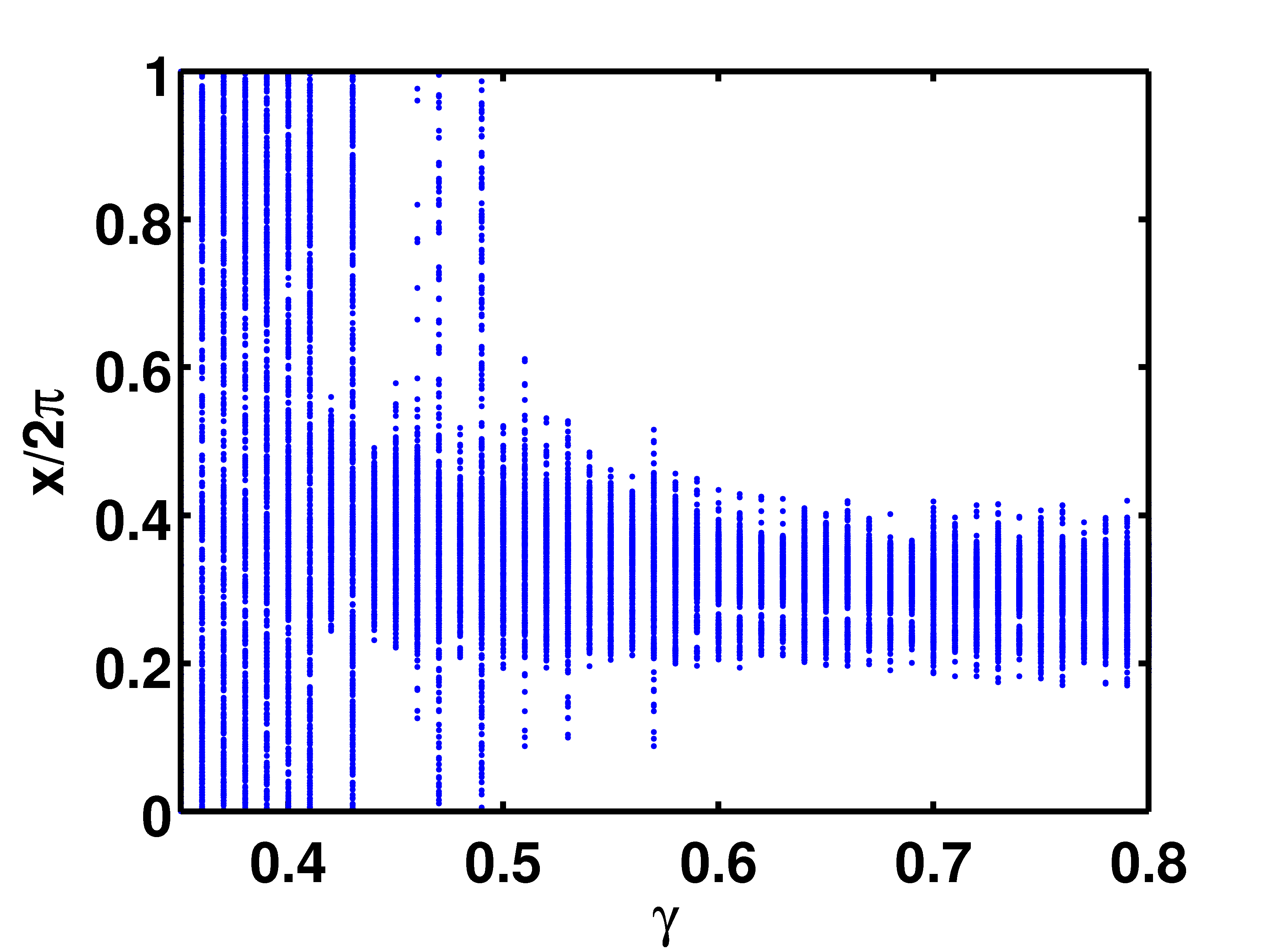}
\caption{\footnotesize\label{fig:Bifurcation_bubble} (Color online) The bifurcation diagram for the embedded ABC map in the bubble regime for the parameter values (A,B,C) =(2.0,1.3,0.05) and $\alpha = 1.1$.}
\end{figure*}

\begin{figure*}
\begin{center}
\begin{tabular}{cc}
 \includegraphics[height=7cm,width=8.5cm]{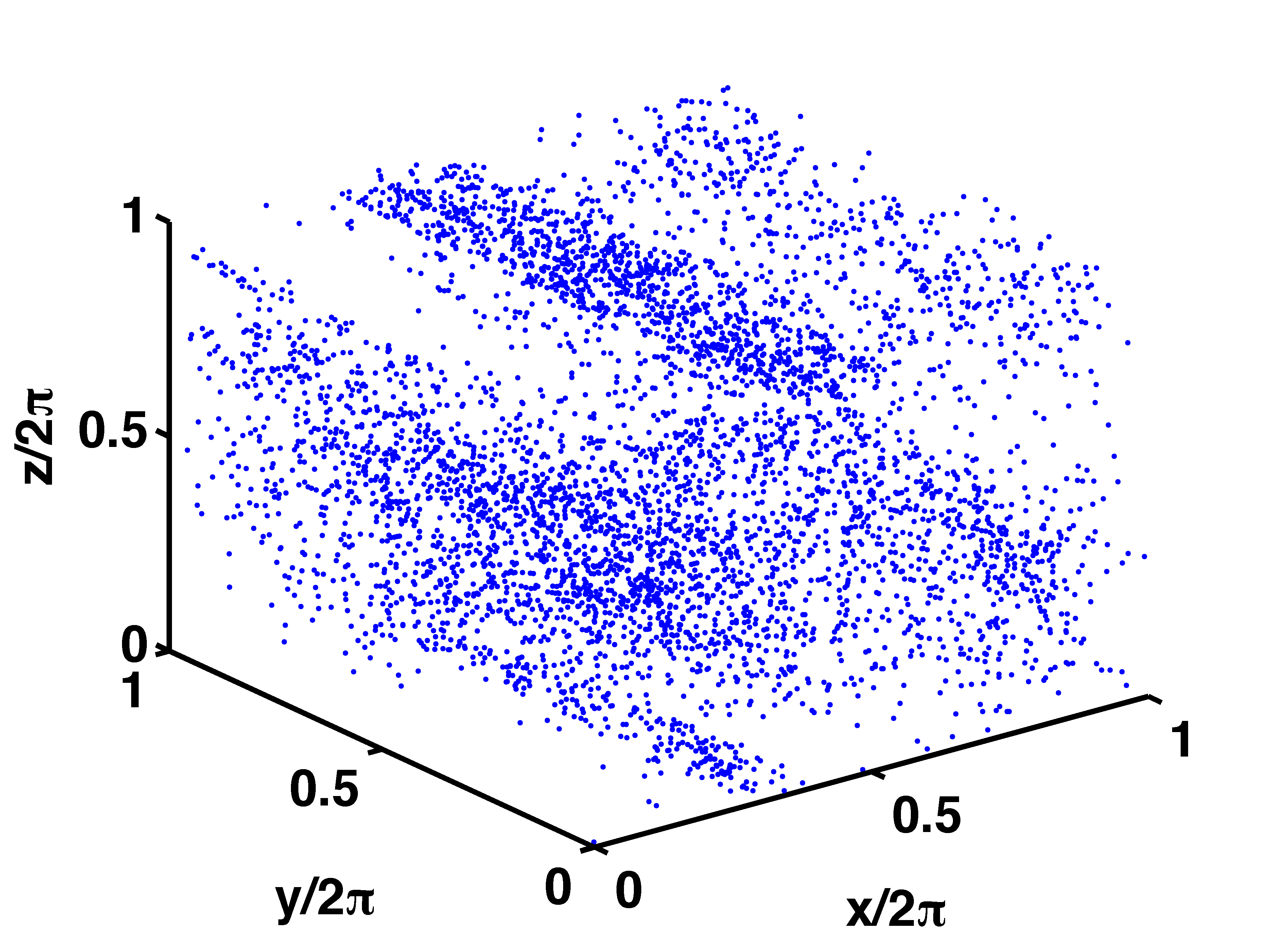}&
\includegraphics[height=7cm,width=8.5cm]{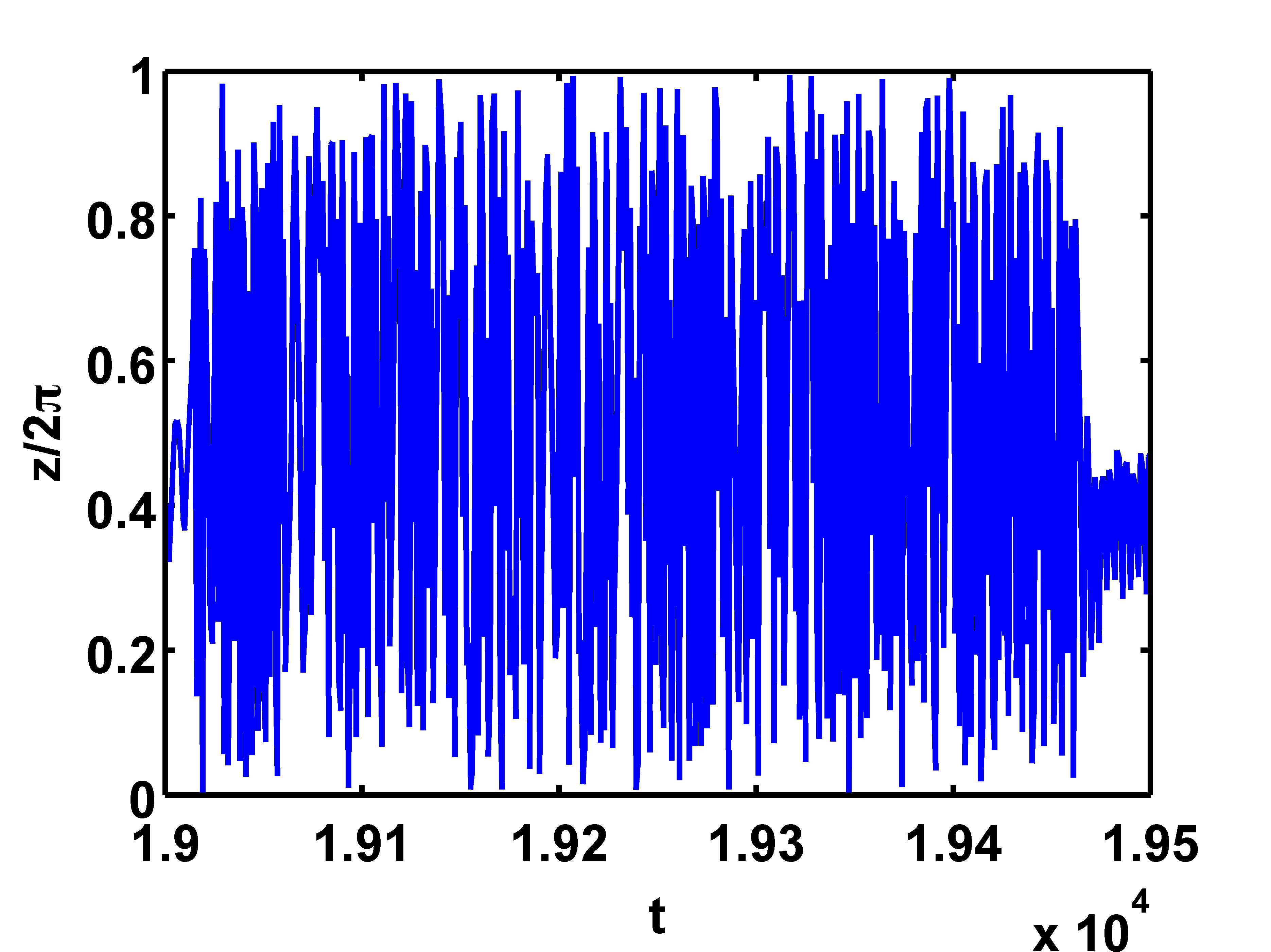}\\
(a) &(b) \\
\end{tabular}{}
\caption{\footnotesize\label{fig:Pre_crisis_Bubble} (Color online) The pre crisis scenario for the embedded two-action ABC map in the bubble regime at the parameter values (A,B,C)=(2.0,1.3,0.05) and $ (\alpha,\gamma) = (1.1, .405)$, (a) the attractor in x-y-z phase space  (500 transients discarded) and (b) the time series.}
\end{center}
\end{figure*}

\begin{figure*}
\begin{center}
\begin{tabular}{cc}
 \includegraphics[height=7cm,width=8.5cm]{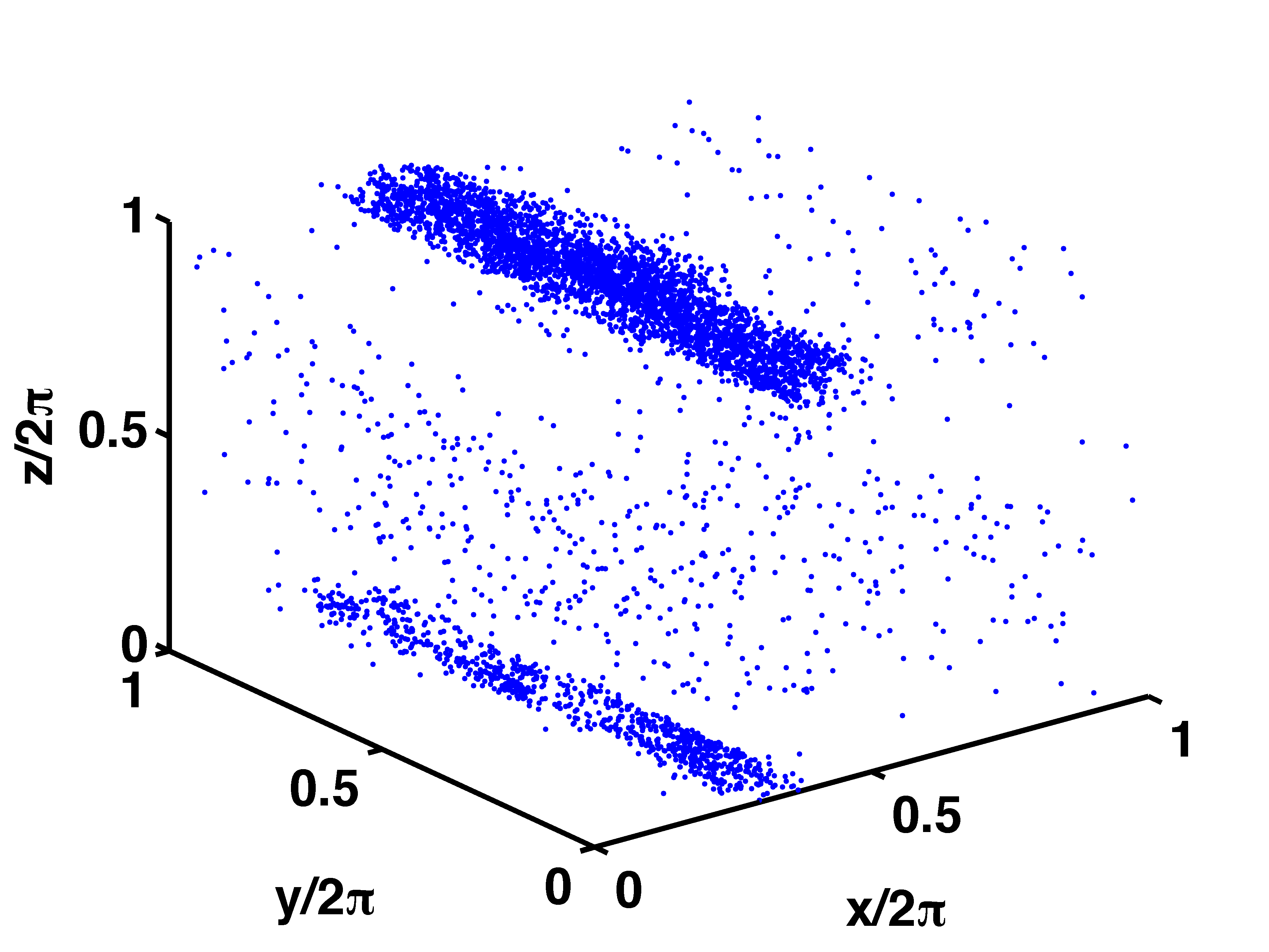}&
\includegraphics[height=7cm,width=8.5cm]{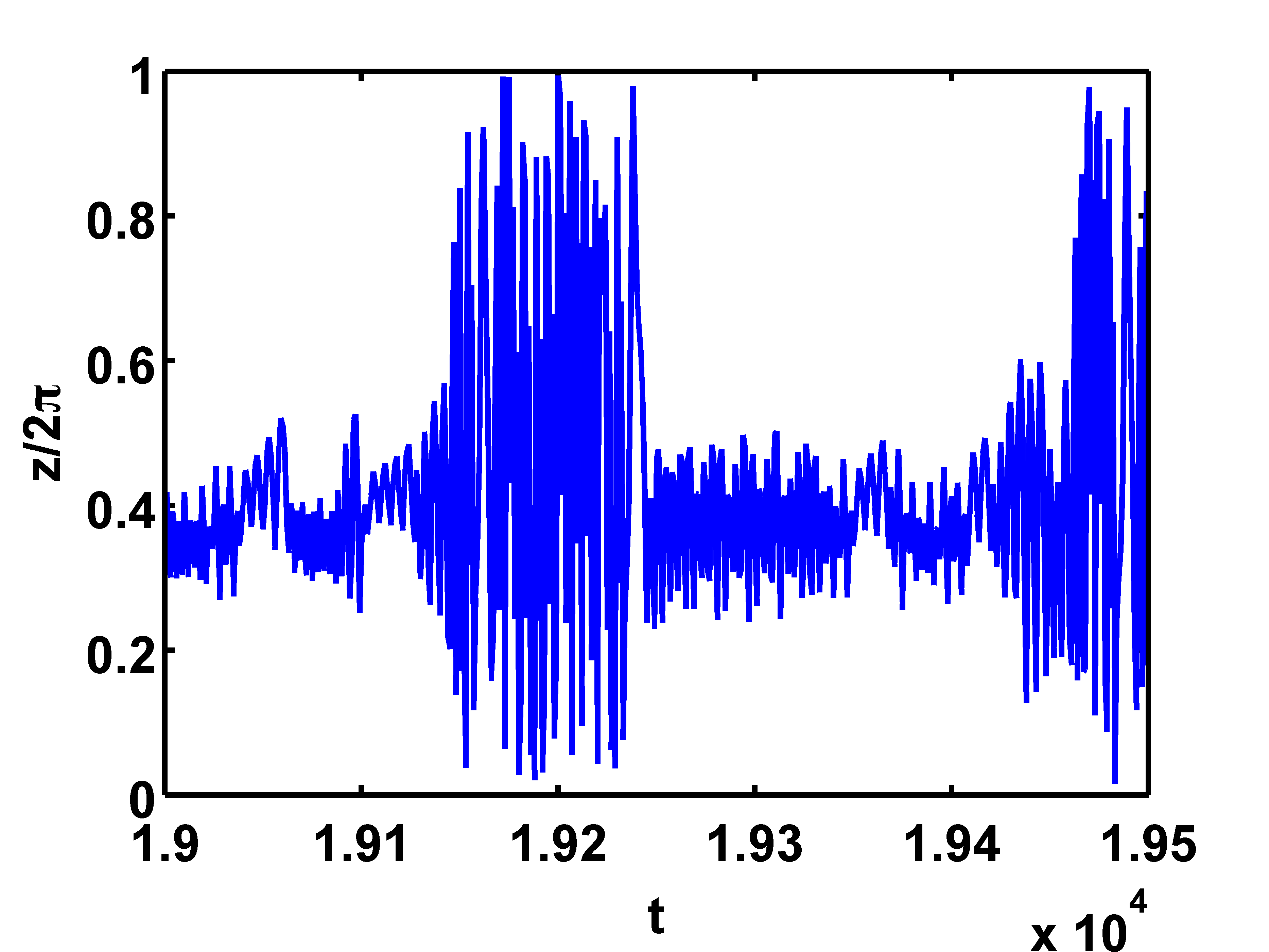}\\
(a) &(b) \\
\end{tabular}{}
\caption{\footnotesize\label{fig:Post_crisis_Bubble} (Color online) The post crisis scenario for the embedded two-action ABC map in the bubble regime for the parameter values (A,B,C)=(2.0,1.3,0.05) and $ (\alpha,\gamma) = (1.1, .425)$, (a) the attractor in x-y-z phase space  (500 transients discarded) and (b) the time series}
\end{center}
\end{figure*}

\begin{figure*}
\begin{center}
\begin{tabular}{cc}
\includegraphics[height=7cm,width=8.5cm]{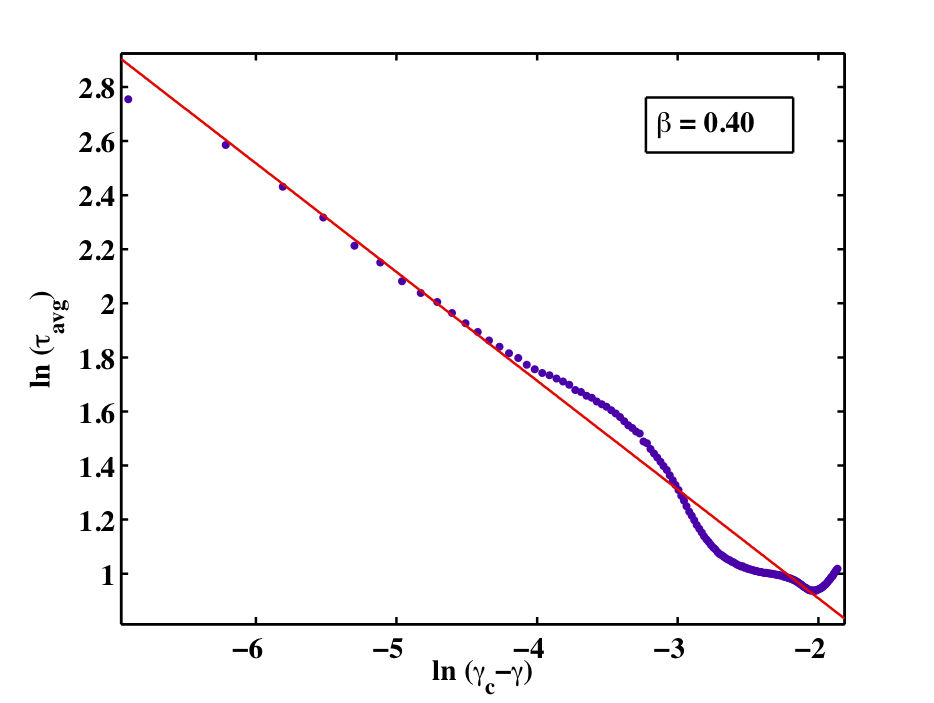}&
\includegraphics[height=7cm,width=8.5cm]{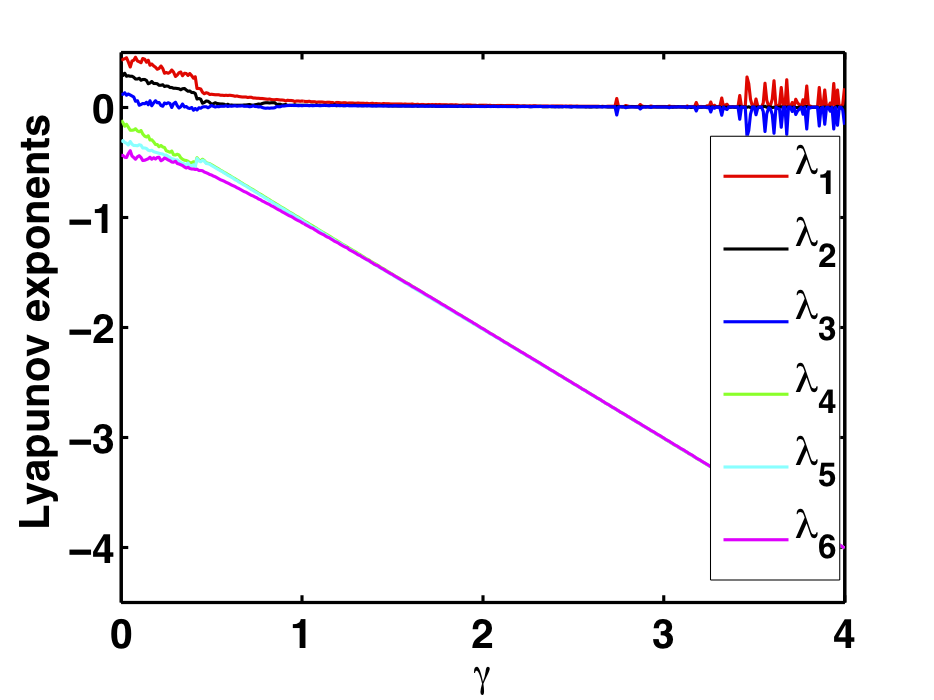}\\(a)&(b) \\
\end{tabular}{}
\caption{\footnotesize\label{fig:Power_Lyap} (Color online) Plots of
phenomena observed in the bubble regime at the parameter values
(A,B,C)=(2.0,1.3,0.05) and $ \alpha = 1.1$, (a) the  power law for  crisis induced intermittency (No. of initial conditions = 500, $\gamma$ varies from 0.26 to  $\gamma_{c} = 0.415$, and (b) the Lyapunov spectrum ($\lambda_{1} > \lambda_{2} >\lambda_{3} >\lambda_{4} >\lambda_{5} >\lambda_{6}$).}
\end{center}
\end{figure*}

We now discuss the bubble regime, where the  density of the particles is less than the density of the fluid.  
The phenomena of crisis and unstable dimension variability seen in the aerosol regime, are also observed in the bubble regime.  The bifurcation diagram for the bubble regime is shown in Fig.~\ref {fig:Bifurcation_bubble} for the set of parameter values 
$(A,B,C) =(2.0,1.3,0.05)$ and $\alpha= 1.1$.
In this case, the phenomenon of crisis induced intermittency is seen. From the bifurcation diagram, we identify the parameter value at which crisis is seen to be  $\gamma_{c} = 0.415$. The pre-crisis and post crisis situations are shown in Fig.~\ref{fig:Pre_crisis_Bubble} and Fig.~\ref{fig:Post_crisis_Bubble}, together with the corresponding time series. We further observe that the average characteristic time $\tau_{avg}$ for the orbit to stay in the pre-crisis attractor before a burst shows a power law of the form,
\begin{equation}
\tau_{avg}\sim  (\gamma_{c} - \gamma)^{-\beta}
\end{equation}
Here, we find that the exponent $\beta = 0.40$. The log-log plot for the scaling behavior is shown in Fig.~\ref{fig:Power_Lyap}(a). Fig.~\ref{fig:Power_Lyap}(b) shows the full Lyapunov spectrum. Here a small jump can also be observed in the variation of the maximum Lyapunov Exponent  for the system [Fig.~\ref{fig:Power_Lyap}(b)]. We also observe that the three largest exponents come close together near the zero level in the window $1.5<\gamma<3$. The other  three also begin to overlap and fall off with almost the same slope as $\gamma$ increases. 

\begin{figure*}[hbtp]
\centering
\includegraphics[height=7cm,width=8.5cm]{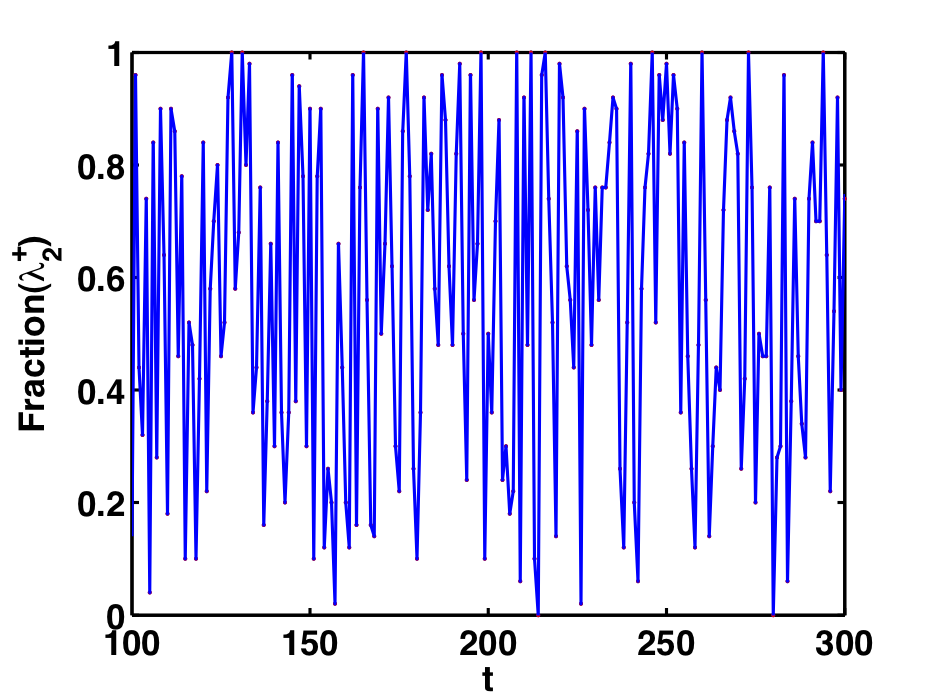}
\caption{\footnotesize\label{fig:UDV_Bubble} (Color online)  Plot of the
number of positive time-50 Lyapunov exponents as a fraction of the total number, versus time,  in the bubble regime for the parameter values (A,B,C)=(2.0,1.3,0.05) and $(\alpha,\gamma) = (1, 2.82)$. The presence of unstable dimension variability is very clear.}
\end{figure*}

Furthermore, the Lyapunov spectrum also shows that the second largest Lyapunov Exponent $\lambda_2$ fluctuates near zero in the window of interest. As in the aerosol case, this signals the presence of unstable  dimension variability (UDV). Calculations of finite time Lyapunov Exponents (FTLE-s) confirm the presence of UDV. We compute time-50 Lyapunov exponents and plot the fraction of positive FTLEs as a function of time. This plot  shows continuous fluctuations [Fig.~\ref{fig:UDV_Bubble}] indicating that the number of expanding and contracting directions is changing at different points in the attractor, a clear signature of unstable dimension variability, as seen before. 

\begin{figure*}[hbtp]
\centering
\includegraphics[height=7cm,width=8.5cm]{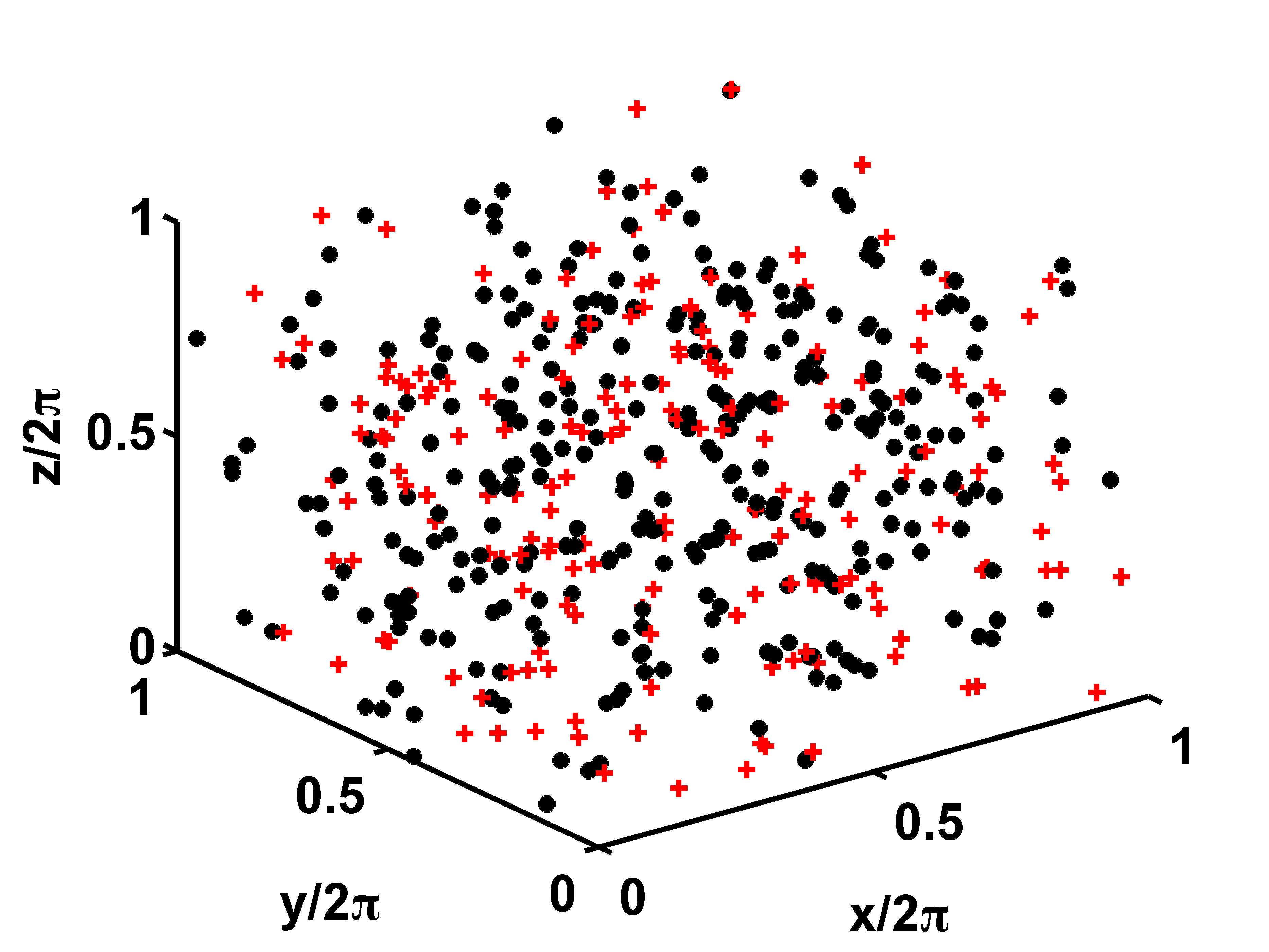}
\caption{\footnotesize\label{fig:Riddled_Bubble} (Color online)  The
riddled basin of attraction in the bubble regime for the parameter
values (A,B,C)=(2.0,1.3,0.05) and $(\alpha,\gamma) = 1, 2.82)$.[Initial
conditions marked with red plus signs '($+$)'  evolve to  the attractor with stripes
and those marked with  black asterisks '($*$)' evolve to the attractor with patches).}
\end{figure*}

\begin{figure*}
\begin{center}
\begin{tabular}{cc}
 \includegraphics[height=7cm,width=8.5cm]{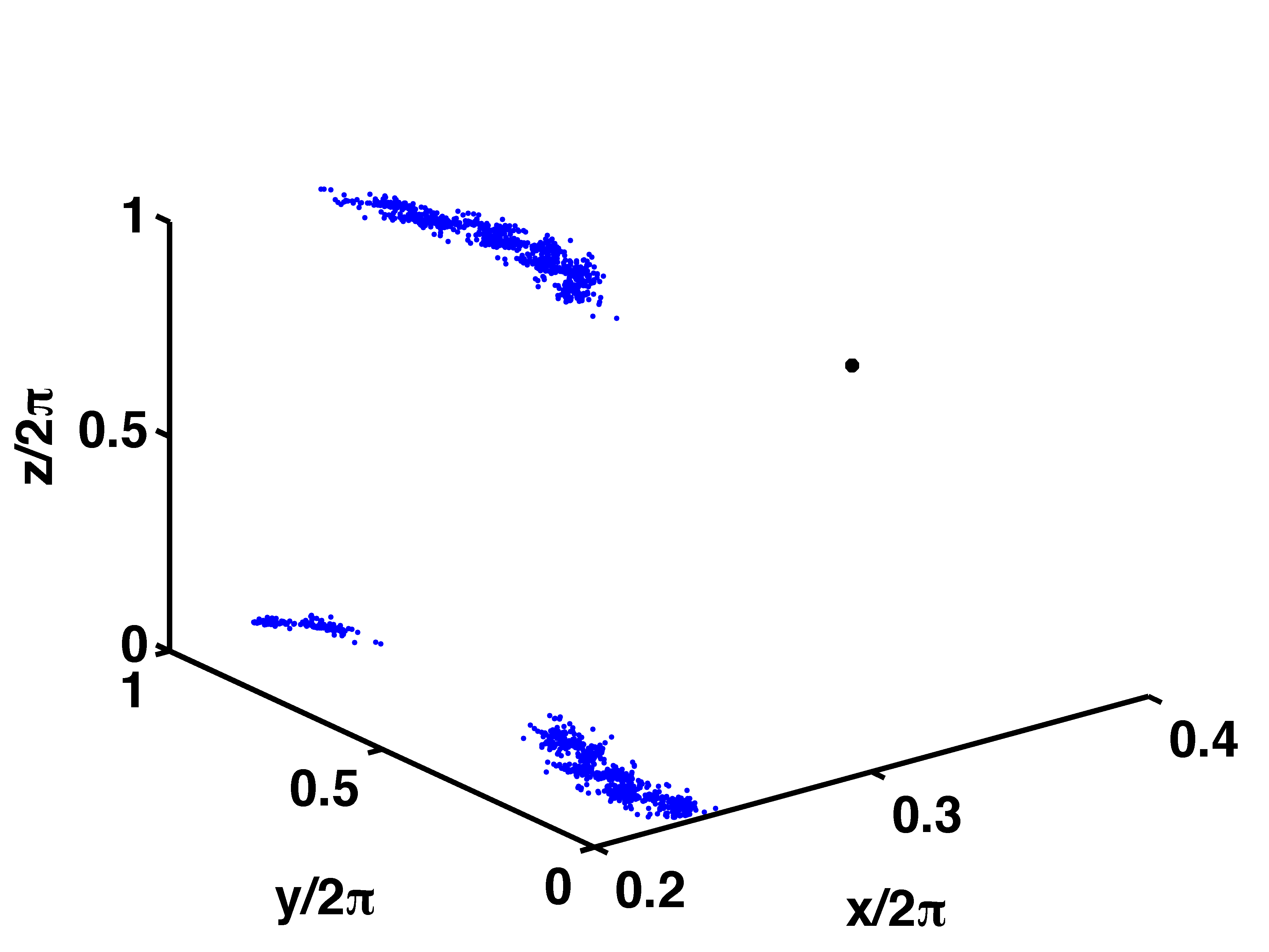}&
\includegraphics[height=7cm,width=8.5cm]{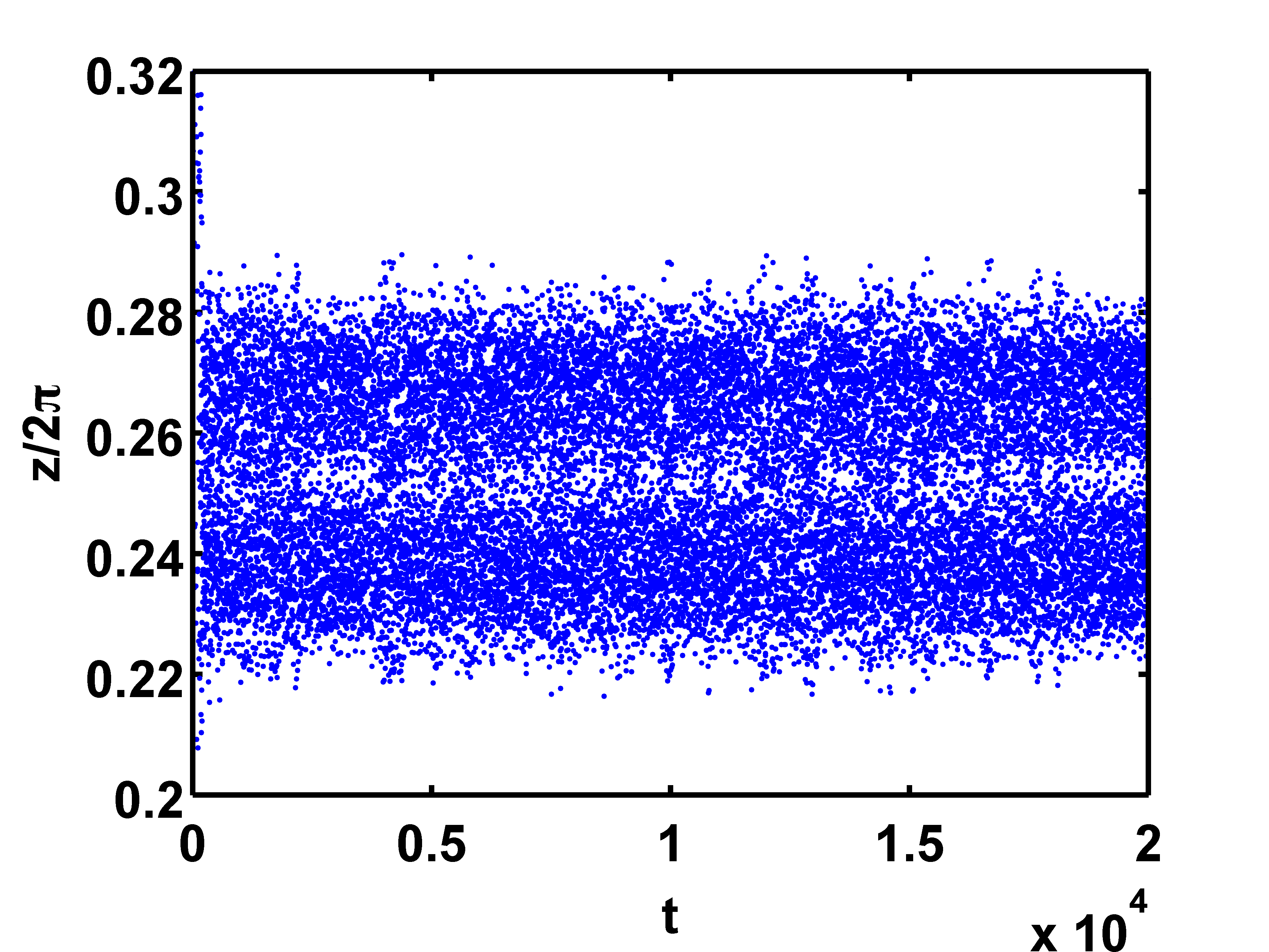}\\
(a) & (b)\\
 \includegraphics[height=7cm,width=8.5cm]{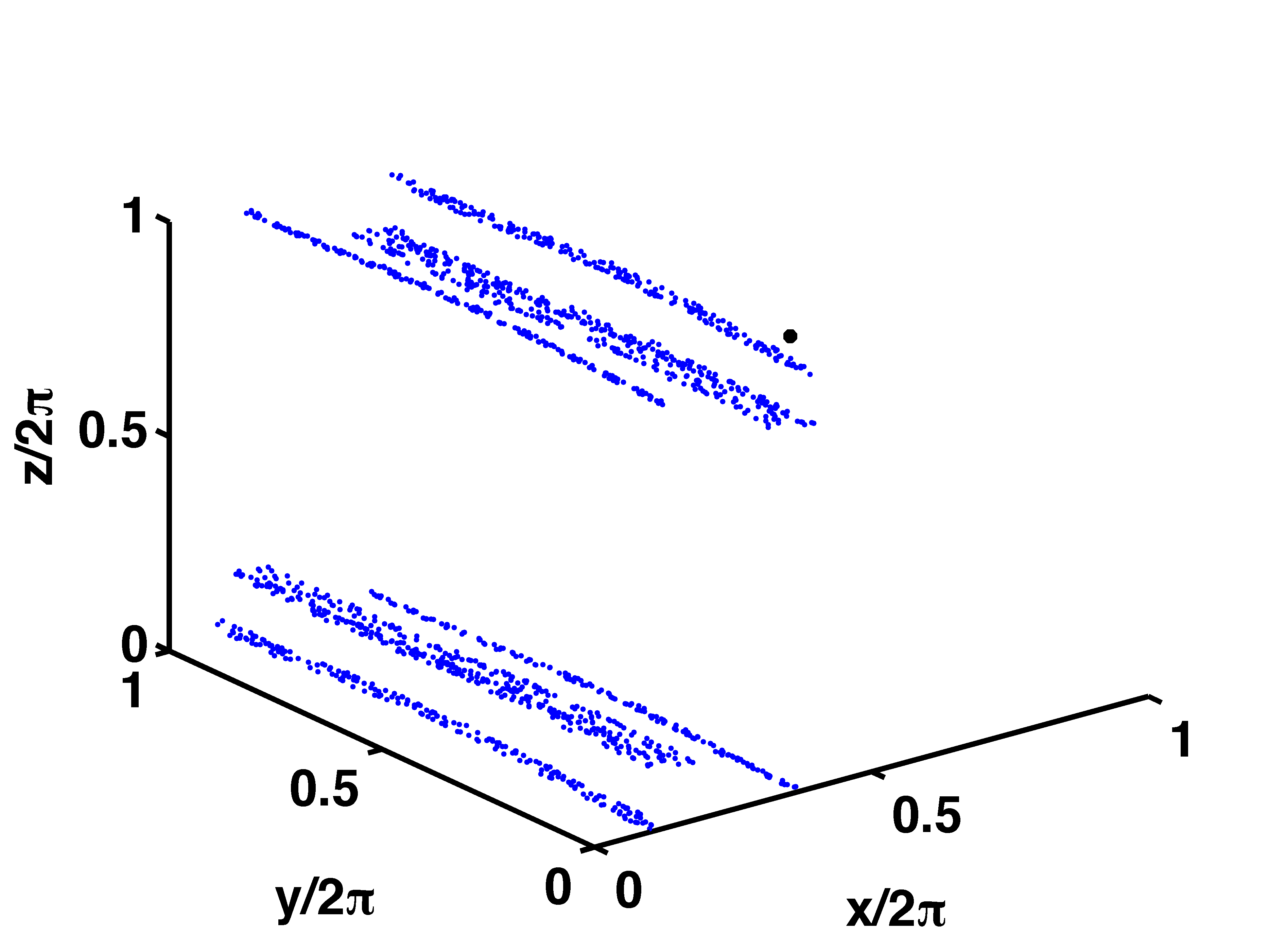}&
\includegraphics[height=7cm,width=8.5cm]{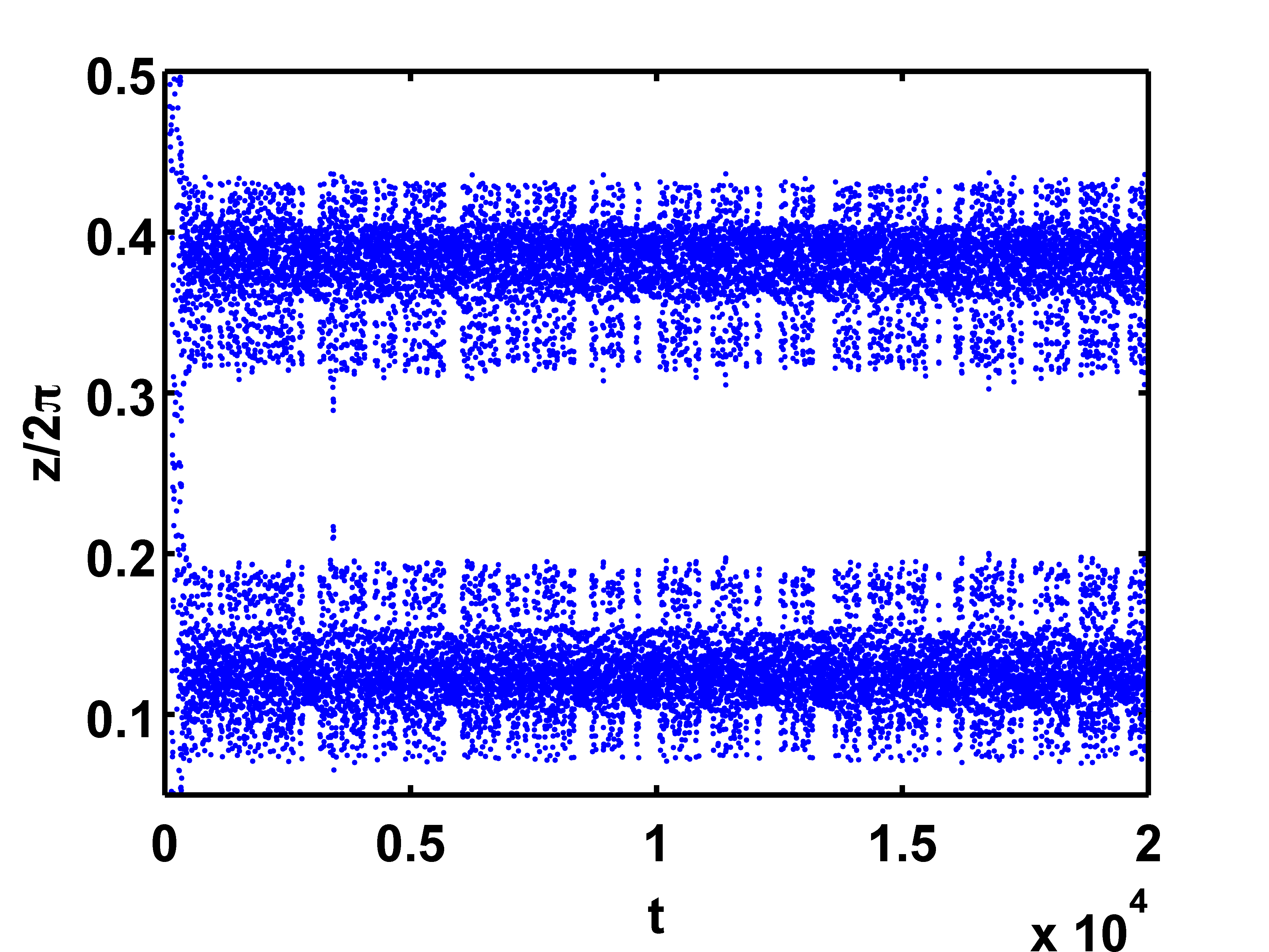}\\
(c) & (d)\\
\end{tabular}{}
\caption{\footnotesize\label{fig:patched_stripped} (Color online)  The riddling of the basin of attraction in the bubble regime for the parameter values (A,B,C)=(2.0,1.3, 0.05)  and $ (\alpha,\gamma) = (1.1, 2.82)$, (a) The attractor with patches (the black dot indicates the initial condition (0.1724, 0.9116, 0.3947)), (b) the corresponding time series points, (c) The attractor with stripes (the black dot indicates the initial condition (0.6003, 0.6661, 0.9543)), and (d) the corresponding time series points.}
\end{center}
\end{figure*}

The basin of attraction of the system in the window $2.5<\gamma<3$ is also of some interest. A given random initial condition asymptotes to the attractor shown in the Fig.~\ref {fig:patched_stripped}(a), where three patches of closely clustered points can be seen. However, another initial condition in an arbitrarily close neighborhood of the previous initial condition asymptotes to a different  attractor [Fig.~\ref{fig:patched_stripped}(c)] with two sets of three stripes each. This is further supported by the time-series of $x$ which shows one-band and two-band structures for nearby initial conditions [Fig.~\ref{fig:patched_stripped}]. The black dots in Fig.~\ref{fig:patched_stripped}(a) and Fig.~\ref{fig:patched_stripped}(c) indicate the positions of the initial conditions (the first 500 iterates are discarded as transients). This clearly indicates the existence of a riddled basin of attraction. The riddled basin of attraction is shown in Fig.~\ref{fig:Riddled_Bubble} for 500 random initial conditions. It may be noted that the number of initial conditions that leads to the attractor with patches is about half of that to the attractor with strips.  Such riddled basins have been seen before in the neighborhood of the crisis in other systems \cite{riddle}, but have not been observed before in the bailout embeddings of maps. We note that in this particular case, the two sets of initial conditions, viz. the conditions that lead to each type of attractor, are randomly distributed in the phase space. This can have interesting consequences for impurity dynamics, including the question of the co-existence of the two attractors. We plan to explore this in further work.

\section{The Phase Diagram of the two-action case}

\begin{figure*}
\begin{center}
\begin{tabular}{cc}
\includegraphics[height=7cm,width=8.5cm]{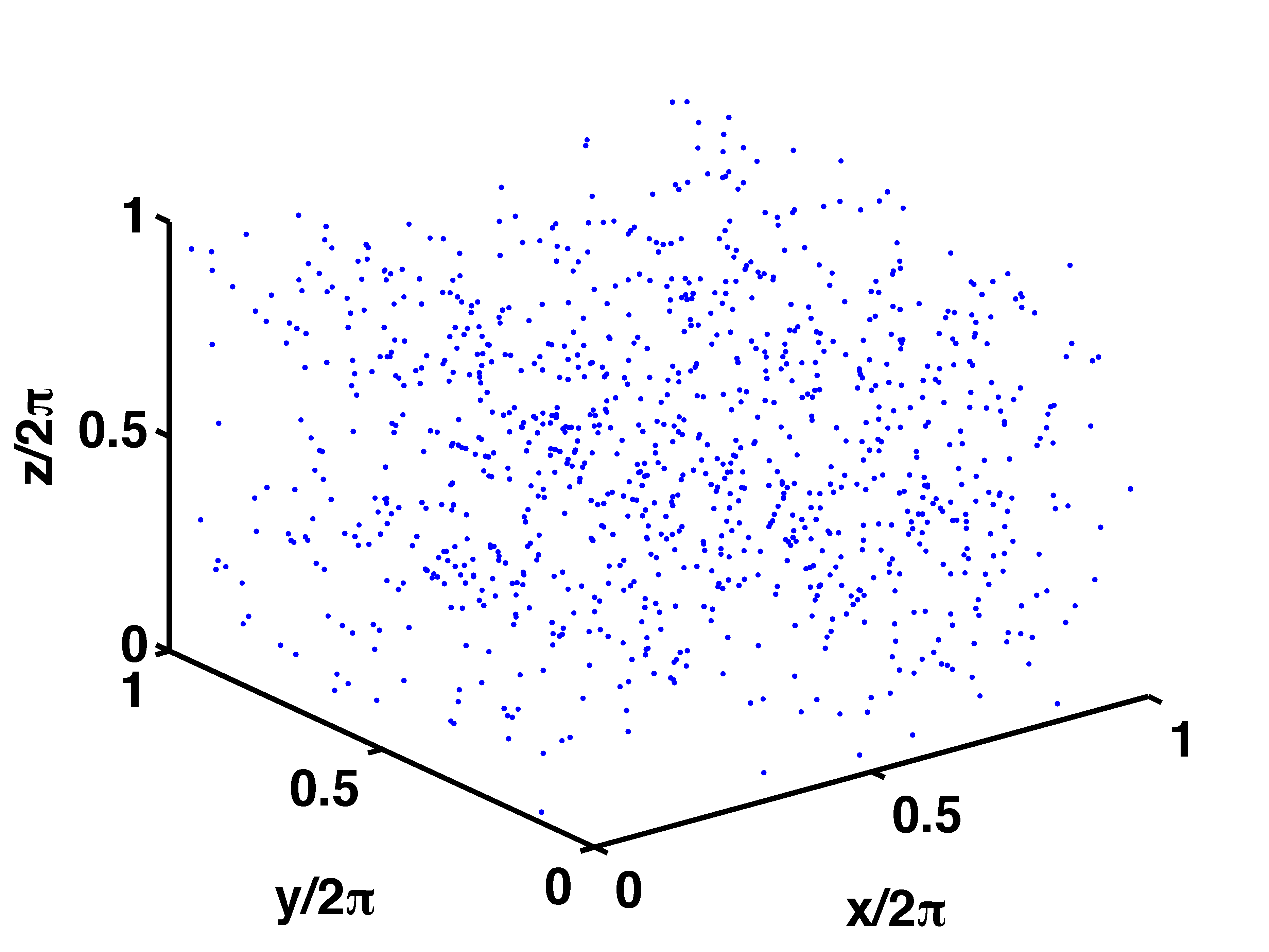}&
\includegraphics[height=7cm,width=8.5cm]{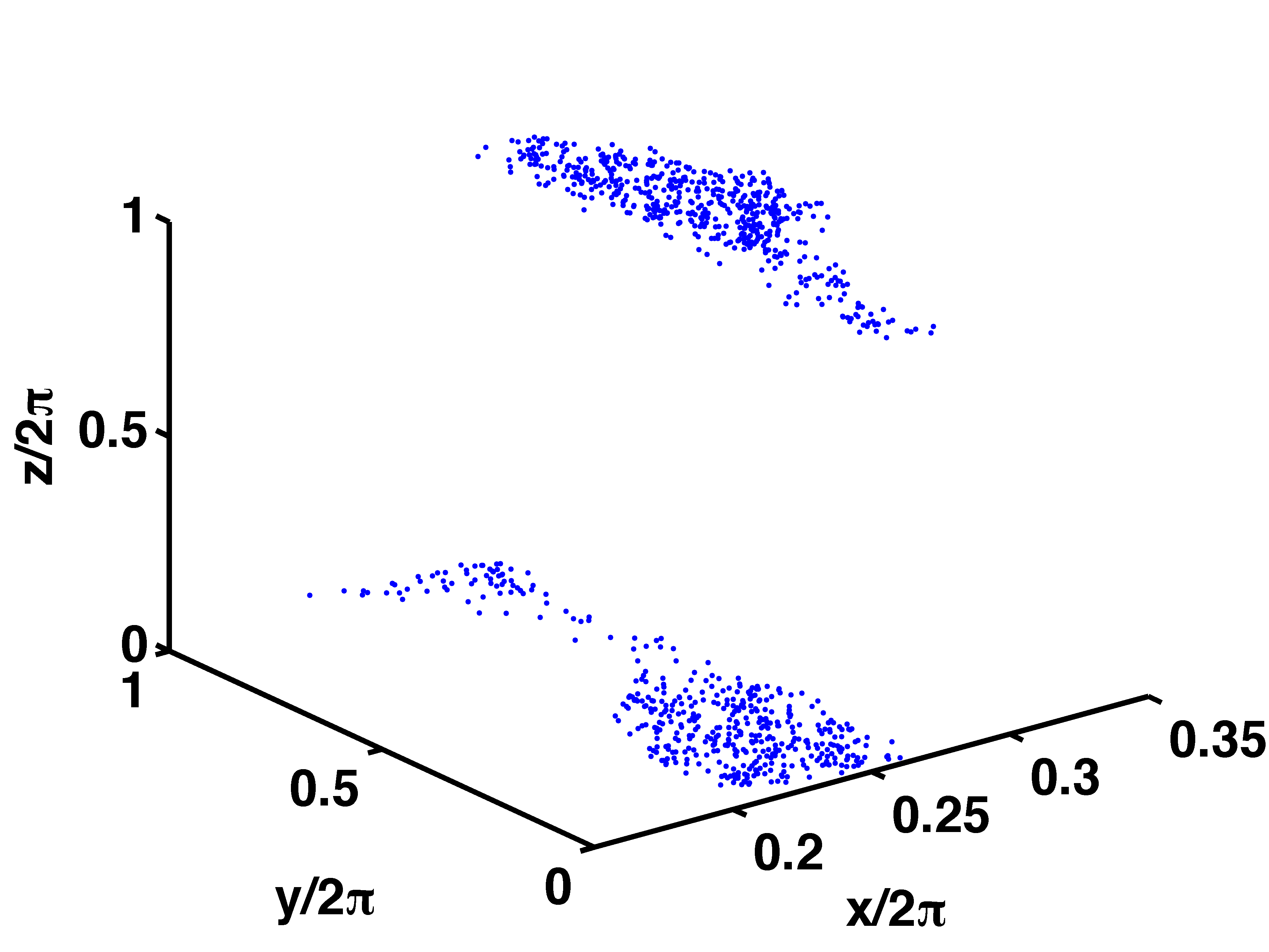}\\(a)&(b) \\
\end{tabular}{}
\caption{\footnotesize\label{fig:Chaotic_Hyperchaotic} (Color online)
Phase space plot  for the chaotic and hyperchaotic regimes for the two-action
for the parameter values (A,B,C)=(2.0,1.3,0.05) and  $ \gamma = 1.5$, (a)
the aerosol regime ($\alpha = 0.7$), and (b) the bubble regime ($\alpha =
1.1$). The last 1000 points of the 10000 iterates are plotted. See Fig.\ref{fig:Phase_diagram} for the complete phase diagram of the two-action case.}
\end{center}
\end{figure*}

\begin{figure*}[hbtp]
\includegraphics[height =7.5cm, width=10cm]{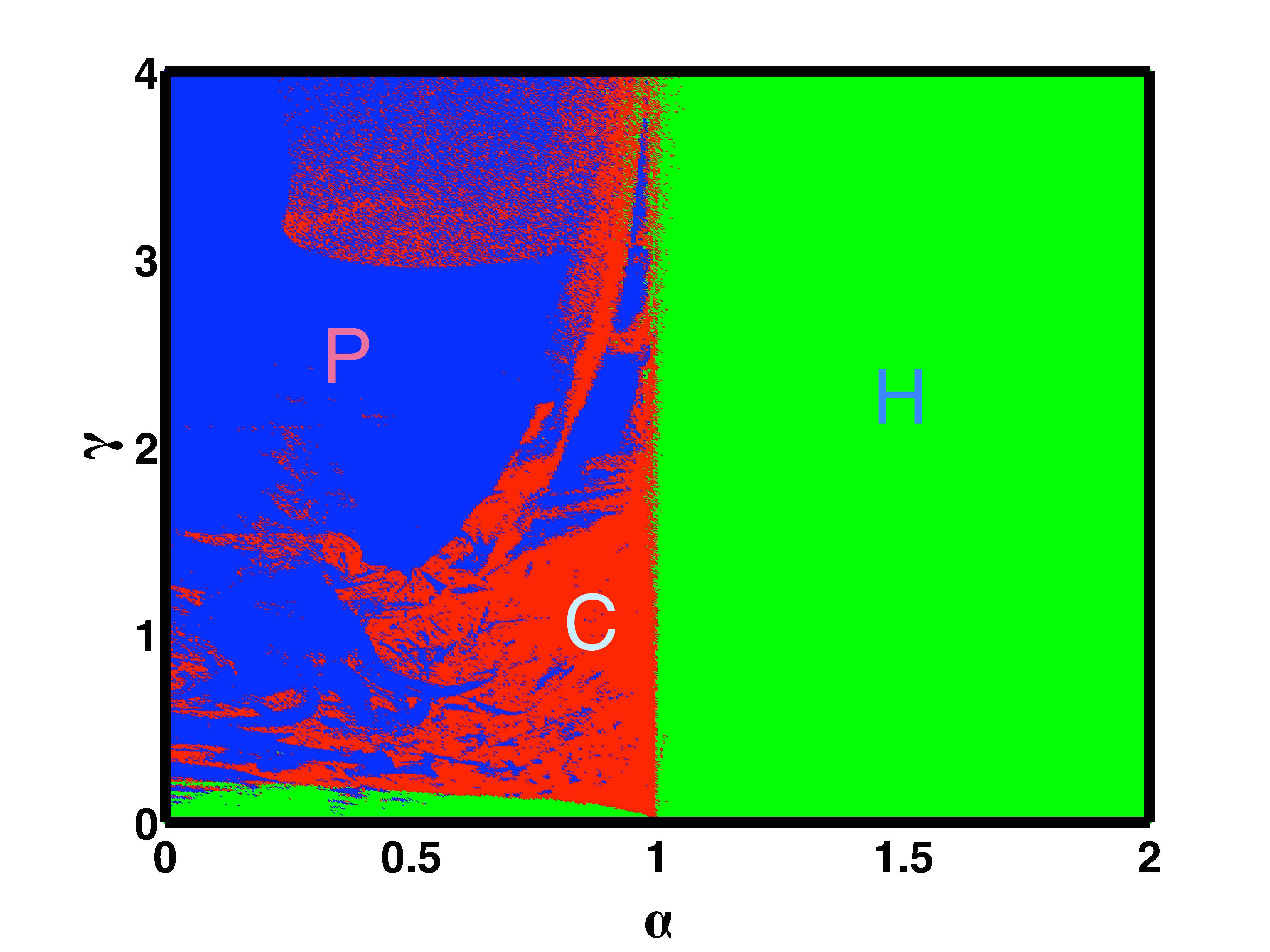}
\caption{\footnotesize \label{fig:Phase_diagram}(Color online) The phase
diagram of the embedded two-action ABC map for parameter values (A,B,C)
= (2.0,1.3,0.05),  (periodic orbits - blue (P), chaotic behavior - red
(C), hyperchaotic regions - green(H)). The $\alpha-\gamma$ space is covered by a $400\times800$ mesh, each of size $0.005\times0.005$. A total of 25000 iterates have been calculated in each case. The phase diagram has been plotted for 15000 iterates discarding first 10000 iterates as transients.  }

\end{figure*}

In order to get the full picture of the dynamical behavior of the
two-action case, we plot the full phase diagram in the parameter space
[Fig.~\ref{fig:Phase_diagram}]. The calculation of all six Lyapunov
exponents of the embedded map in the parameter space helps to visualize
the three kinds of dynamical behavior, viz. periodic orbits, chaotic orbits and
hyperchaotic orbits. We classify the behavior as (a) periodic if
$\lambda_{1},\lambda_{2} <0$, (b) chaotic if $\lambda_{1} >
0,\lambda_{2} \leq0$, and (c) hyperchaotic if $
\lambda_{1}>0,\lambda_{2} >0$. The regimes with $\lambda_{1} >
0,\lambda_{2} \leq0$ shows chaotic behavior as in
Fig.~\ref{fig:Chaotic_Hyperchaotic}(a) whereas the regimes with both
$\lambda_{1} > 0,\lambda_{2} > 0$ shows  hyperchaotic behavior
(Fig.~\ref{fig:Chaotic_Hyperchaotic} (b)) due to the presence of two
diverging directions which may result in a higher efficiency of mixing
and transport in the fluid flow. The phase digram is shown in
Fig.~\ref{fig:Phase_diagram} in which the blue (P), red (C) and green
(H) regions indicate periodic, chaotic and hyperchaotic behavior in the
system, respectively. The aerosol region shows periodic and chaotic
and hyperchaotic behaviour, whereas the bubble region is completely 
hyperchaotic. We note that the phase diagram is plotted after a very
long asymptote, and other kinds of behaviour are seen in the
bubble region in the transient.

We compare our results with earlier work \cite{Nirmal} in which the dynamics of inertial particles on the surface of a fluid is modeled by four dimensional dissipative bailout embedding maps. Here, the $2-d$ standard map is used as the  base flow, leading to the following map equations for the embedded map \cite{Nirmal}. 
\begin{empheq}[right=\empheqrbrace \mod 2\pi]{align}
x_{n+1} = x_{n}+\frac{K}{2\pi}\sin(2\pi y_{n})+\delta^{x}_{n} \nonumber\\
y_{n+1} = y_{n}+\frac{K}{2\pi}\sin(2\pi y_{n})+\delta^{y}_{n} \nonumber\\
\delta_{n+1}^{x} = e^{-\gamma}[\alpha x_{n+1}-(x_{n+1} - \delta_{n}^{x}) \nonumber]\\
\delta_{n+1}^{y} = e^{-\gamma}[\alpha y_{n+1}-(y_{n+1} - \delta_{n}^{y})]
\end{empheq}
The parameter $K$ controls the chaoticity of the map and $\alpha$ and
$\gamma$ are the usual mass ratio and dissipation parameters respectively.
This map exhibits crisis induced intermittency in the aerosol regime
with characteristic times $\tau$ between the bursts which show power-law
scaling, where  $\tau \approx (\gamma_{c}-\gamma)^{-\beta}$ where the exponent $\beta =
0.35$, and $\gamma_{c}$ is the critical value at which the crisis
occurs. Moreover, the crisis seen here is of the interior type similar
to what we find in the present study for the $3-d$ case.  However, in
contrast, the embedded $ABC$ map we have studied does not show
intermittency and associated power law behavior in the aerosol regime.
Prior to the study of Ref. \cite{Nirmal}, it was believed that only
bubbles are capable of breaching the  elliptical islands in the phase
space and targeting  periodic structures.  However, both  the study of
the $2-d$ standard map carried out in Ref. \cite{Nirmal}, as well as the  $3-d$ ABC map studied here show  that such dynamics is demonstrated by aerosols as well, at certain values of the dissipation parameter. Therefore, it is clear  that the dissipation parameter $\gamma$ has a crucial role to play in  the preferential concentration of inertial particles and in  the targeting of  periodic structures. 

The bubble regimes of the embedded  Standard map and the embedded $ABC$ map also differ in crucial ways.
No crisis is seen in the bubble regime of the embedded Standard Map.  In contrast, the  $3-d$  ABC map case shows  crisis induced intermittency in the bubble regime with power law behavior as a function of $\gamma -\gamma_c$ in the characteristic times between bursts. It may be noted that this situation is similar to that seen in the  aerosol regime in the $2-d$ standard map case. In addition, we also observe that the riddled basin and UDV are seen in the bubble regime in the present $3-d$ ABC map case, were not seen in the Standard map case. 

The phase diagram for the embedded Standard map shows periodic
behavior,  chaotic structures,  as well as mixing (\cite{FN}) in the
bubble regime\cite{Nirmal}. The phase diagram  of the embedded $ABC$ map,
(Fig.~\ref{fig:Phase_diagram}) shows fully hyperchaotic regions. Prominent tongues corresponding to periodic regimes  are seen in the phase diagram  for the $2-d$ standard map case but are absent in the 3D case. The phase diagram of the $2-d$ case is characterized by sharp  boundaries, whereas the boundaries in the $3-d$ case are not sharply defined.

In the case of the embedded Standard map, it was shown that the transport and diffusive properties of the system were strongly correlated with the phase diagram.
Here, for the $3-d$ embedded $ABC$ map as well, the most important effects of dimensionality are expected to be seen in the transport properties where the detailed structure of the phase diagram is expected to influence diffusivity. We plan to explore the transport and diffusivity of the embedded ABC map, explore the effects of the barriers in phase space, and compare these with the phase diagram and the effects seen in the  $2-d$ case in subsequent work. 

As mentioned in the introduction, the one action case of the ABC map shows
transport properties which are quite distinct from the two action case. Given the relation between transport and dynamics seen in the case of the Standard map, we carry out a brief study of the dynamical behaviors of the 
one action case of the embedded $ABC$ map in the next section. 

\section{The one-action case}
\begin{figure*}
\begin{tabular}{cc}
 \includegraphics[height=7cm,width=8.5cm]{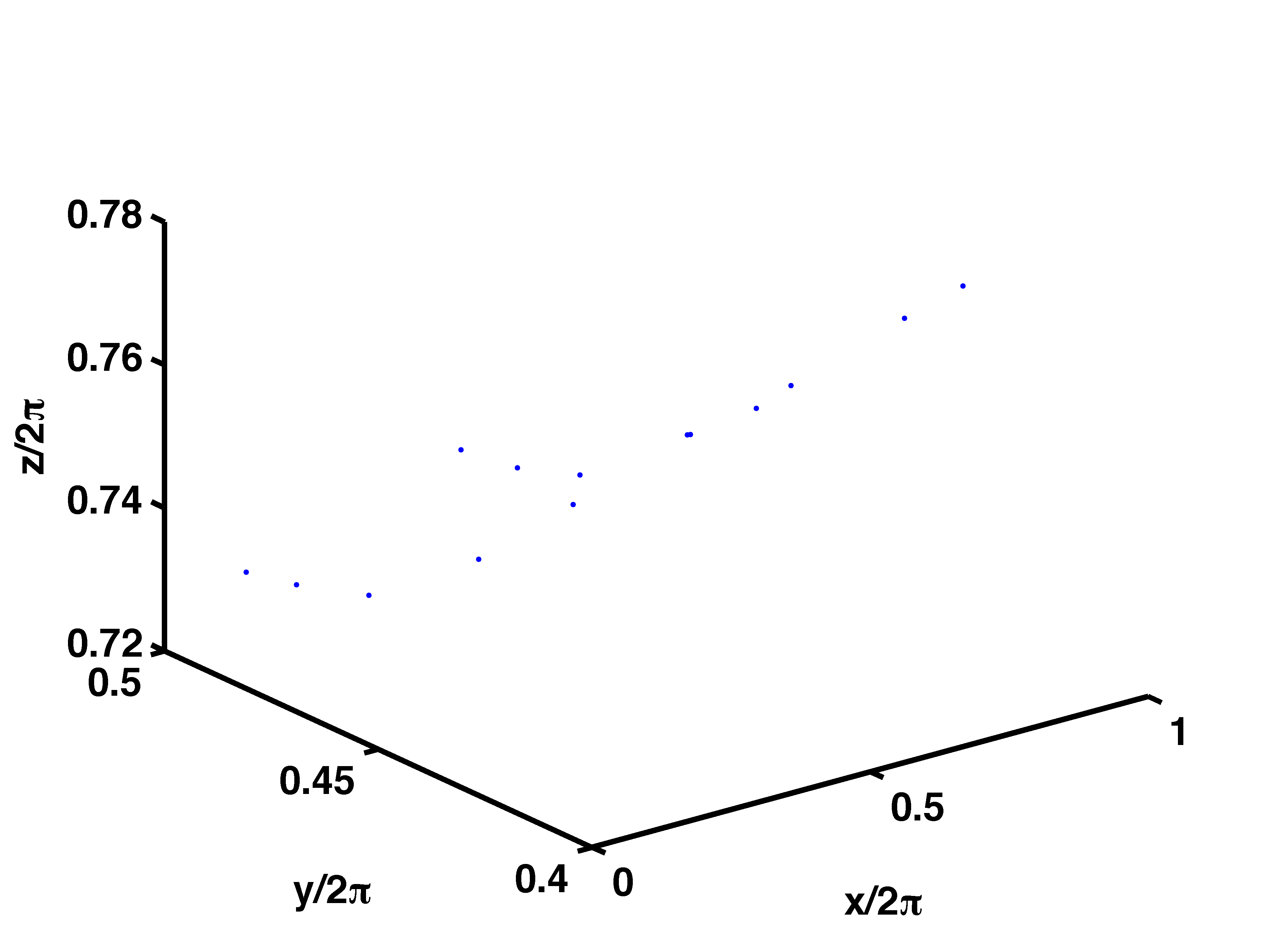}&
\includegraphics[height=7cm,width=8.5cm]{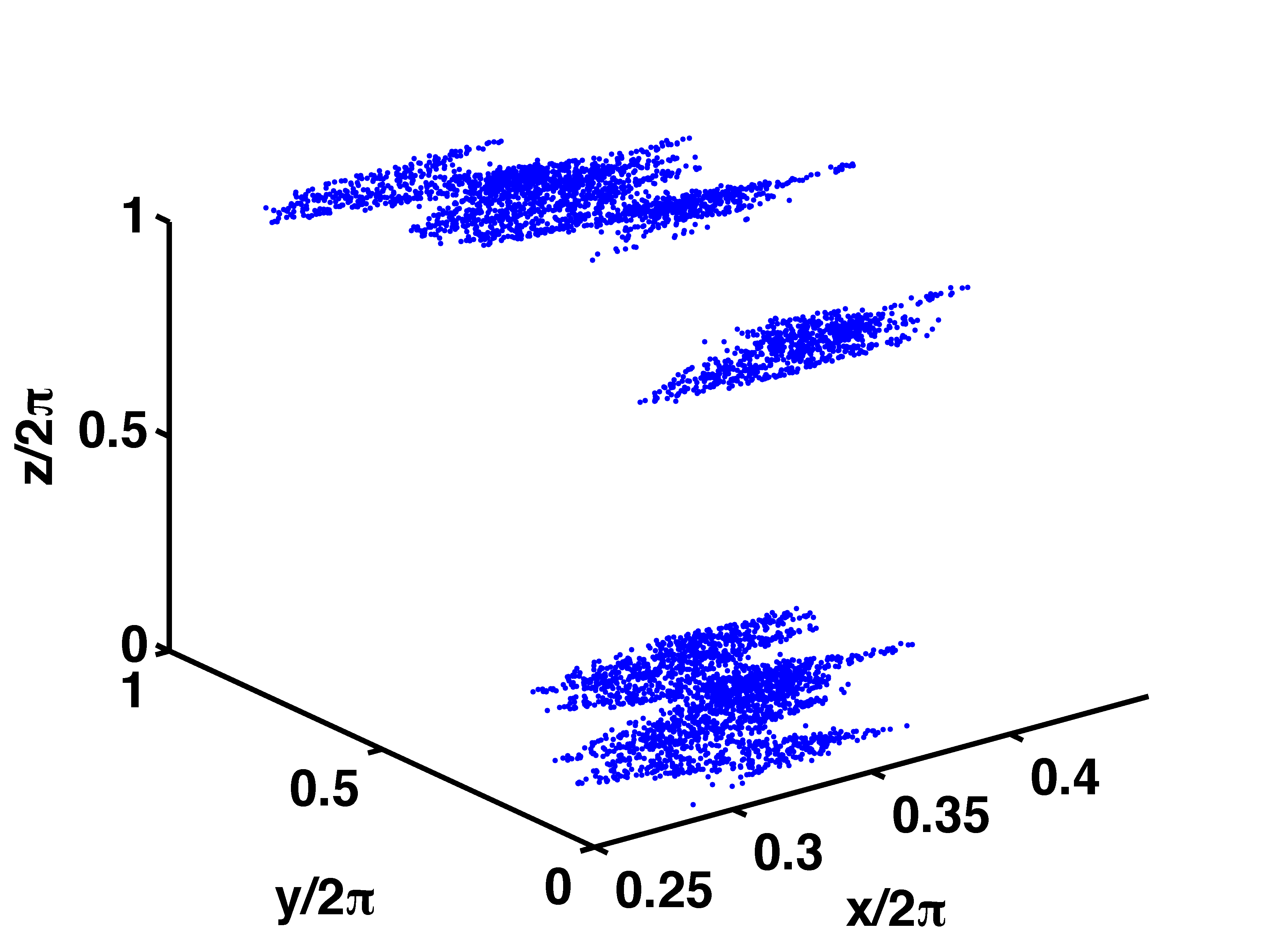}\\
(a) & (b)
\end{tabular}{}
\caption{\label{fig:embed_one_ABC} \footnotesize (Color online) The x-y-z phase space of the embedded one-action ABC map for parameter values (A,B,C)=(1.5,0.08,0.16) in (a) the aerosol regime for $ (\alpha,\gamma) = (0.2,2.9)$, and (b) the bubble regime for $ (\alpha,\gamma) = (1.2, 2.8).$}
\end{figure*}

Here, we study the one-action embedded ABC map at the parameter values $(A,B,C) = (1.5,0.08,0.16)$ and examine both the aerosol and the bubble regimes. Fig.~\ref{fig:embed_one_ABC} shows the phase space portraits of the system at $\alpha = 0.2$ (in Fig.~\ref{fig:embed_one_ABC}(a)) and $\alpha= 1.2$ (in Fig.~\ref{fig:embed_one_ABC}(b)). We find periodic behavior in the aerosol regime, whereas the bubble regime shows hyperchaotic  regions as discussed below.

\subsection{The aerosol regime: Periodic behavior}

The most striking feature observed here, for the one-action map in the aerosol regime is that for the parameter values $(A,B,C)=(1.5,0.08, 0.16)$, $\alpha =0.2$ and $\gamma \geq 0.8$, the trajectories evolve rapidly towards  attractors where all points belong to fixed periods. We note that distinct initial conditions settle  onto distinct trajectories, however several distinct trajectories have the same period. This periodic behavior is sensitive to the values of the inertial and dissipation parameters, $\alpha$ and $\gamma$ respectively. With increase in the value of  $\alpha$, we find that the periodicity of points in a given attractor also increases. Also, the number of distinct periodic trajectories of the same period increases with $\gamma$. Interestingly, all the periods observed here, have been even. 

\begin{figure*}
\begin{tabular}{cc}
 \includegraphics[height=7cm,width=8.5cm]{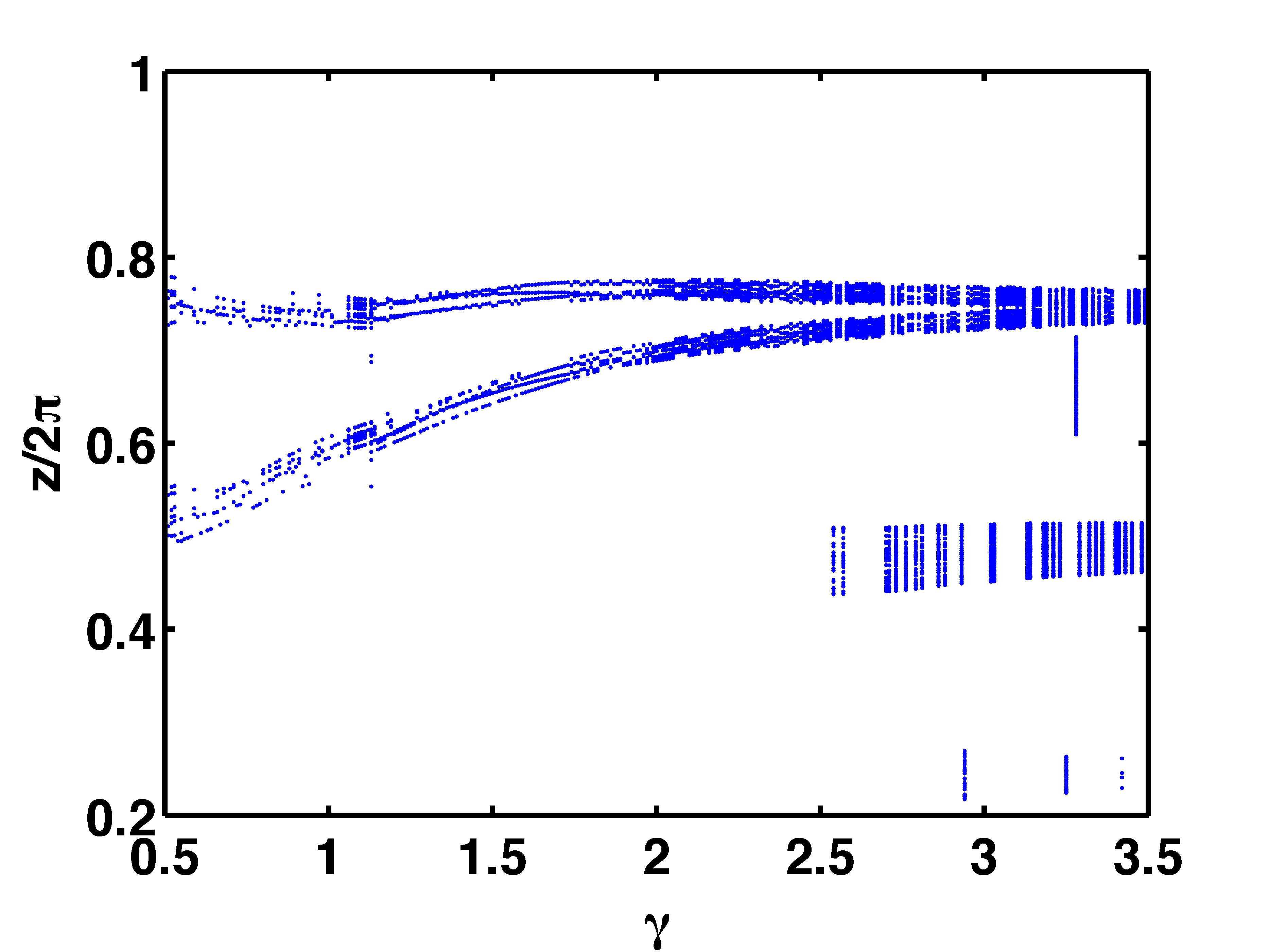}&
\includegraphics[height=7cm,width=8.5cm]{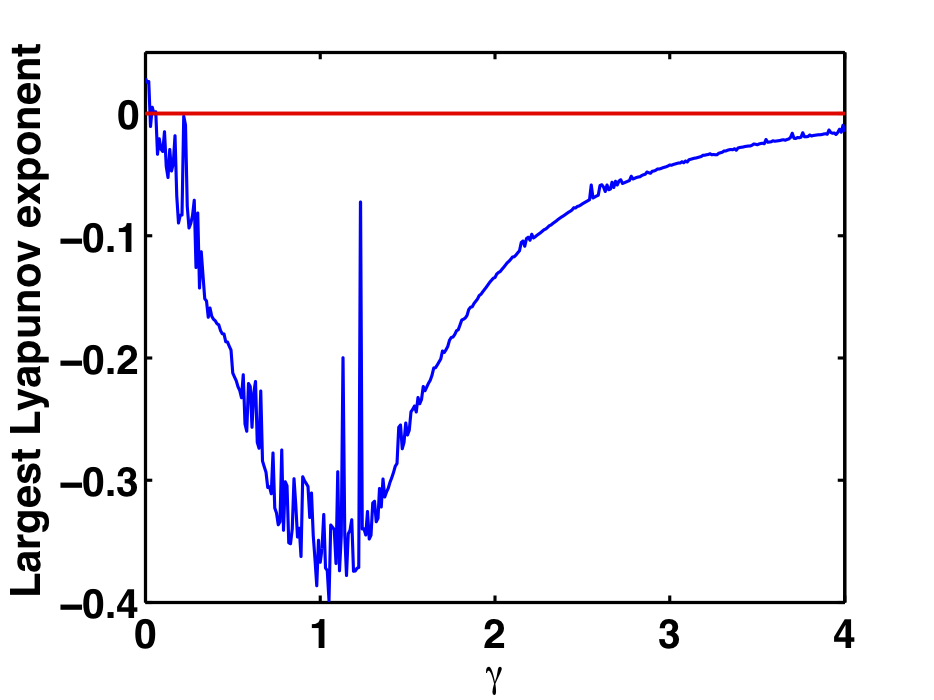}\\
(a) & (b)
\end{tabular}{}
\caption{\label{fig:one_action_aero} \footnotesize (Color online) One-action map case (A,B,C) = 1.5,0.08,0.16) and $\alpha$ = 0.2. (a) the bifurcation diagram, and (b) the Largest Lyapunov exponent variation.}
\end{figure*}

The $x-y-z $ phase space of the system for parameter values (A,B,C) = (1.5,0.08,0.16) and ($\alpha,\gamma$) = (0.2,2.9) is shown in Fig.~\ref{fig:embed_one_ABC}(a). We observe five different periodic attractors in this case, with periods $14, 26, 50, 104, 150$.  We also plot the bifurcation diagram (z vs.$\gamma$) in Fig.~\ref{fig:one_action_aero}(a) which shows windows of crisis for $\gamma > 2.5$ for parameter values $(A,B,C) = (1.5,0.08,0.16)$ and $\alpha$ = 0.2. Furthermore, the behavior of the largest Lyapunov exponent in Fig.~\ref{fig:one_action_aero}(b) confirms the existence of regular orbits for almost the entire range  $0 < \gamma < 4$. 

\subsection{The bubble regime: Crisis and hyperchaotic regions}
The phenomenon of crisis is  also visible in the bubble regime of the one action case. The bifurcation diagram identifies the $\gamma$ value at  crisis to be  $\gamma_c=2.59$, where we anticipate a sudden change in the size of the attractor. However, in contrast to the two-action bubble regime (see section IV.B), we do not see any intermittent behavior induced by crisis. 
The bubble regimes are found to be highly hyperchaotic.  Here, hyperchaos is observed for $0<\gamma\leq4$. We plan to explore the full phase diagram of the system in future work, and also to carry out diffusion and transport studies, and correlate the two. It is expected that the one action and two action cases will show significant differences for the dynamics and diffusion studies. 

\begin{figure*}
\begin{tabular}{cc}
 \includegraphics[height=7cm,width=8.5cm]{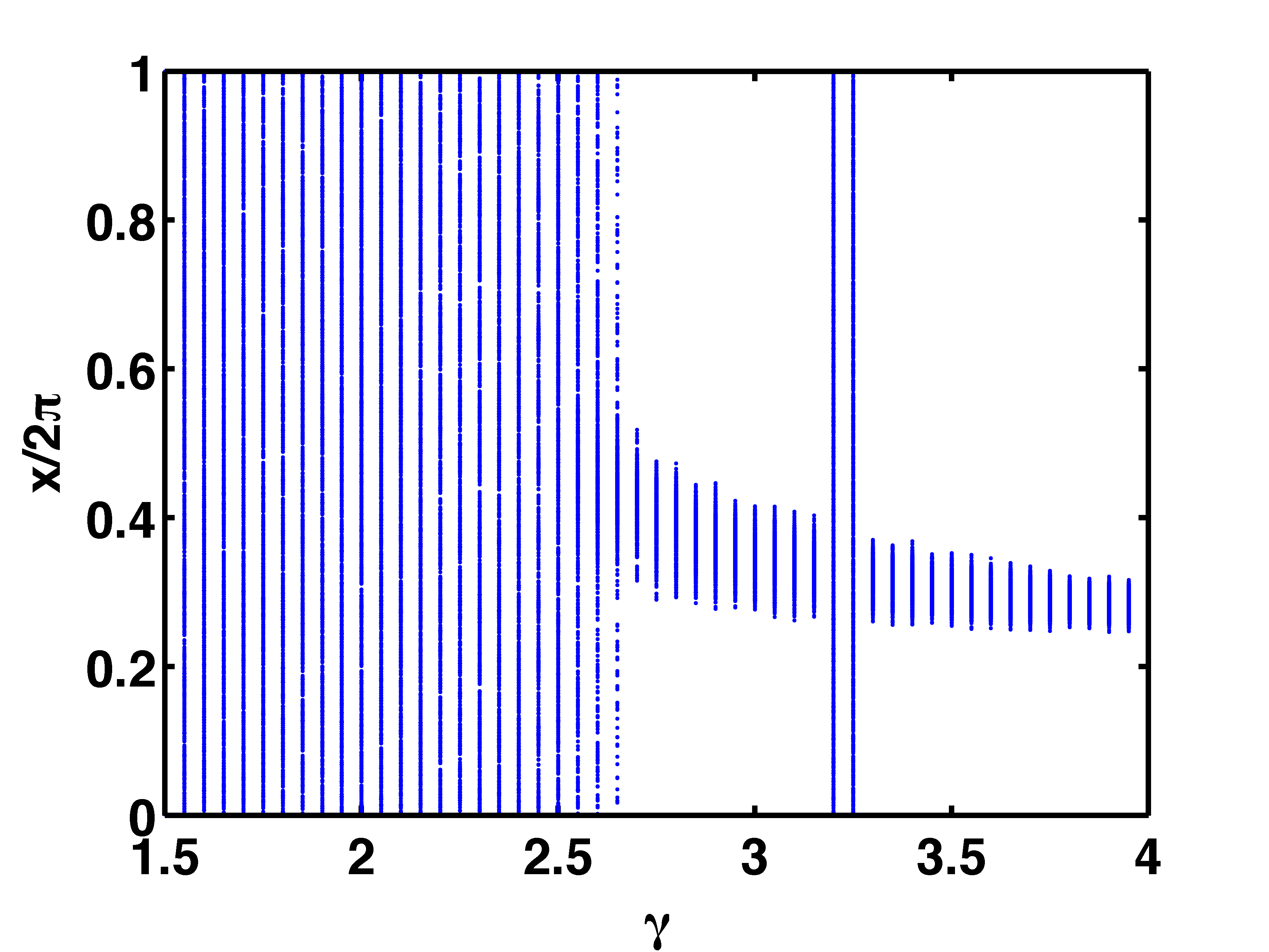}&
\includegraphics[height=7cm,width=8.5cm]{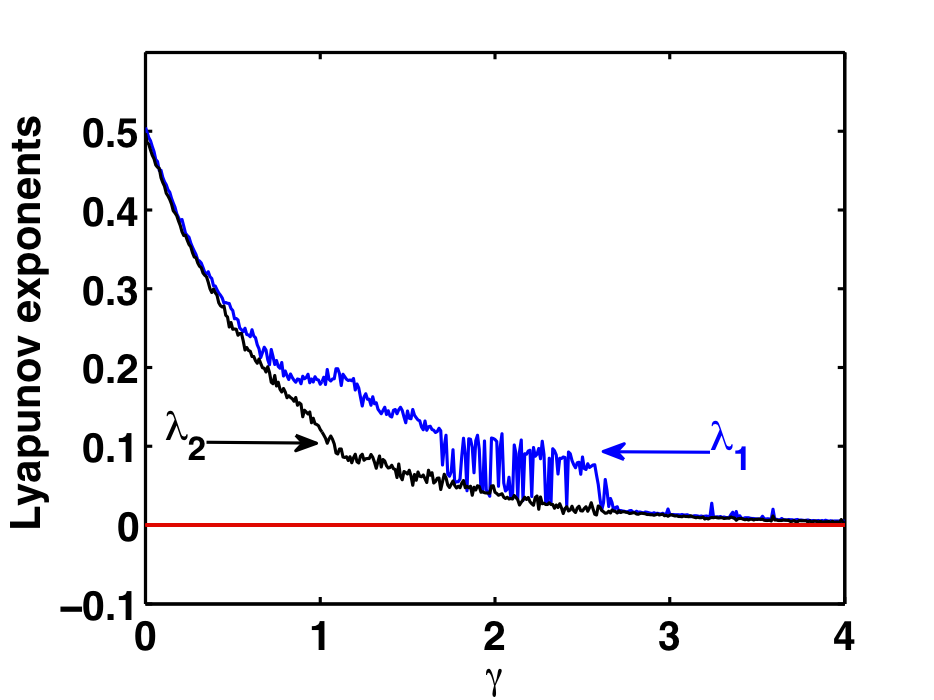}\\
(a) & (b)
\end{tabular}{}
\caption{\label{fig:one_action_bubble} \footnotesize (Color online) One-action map case (A,B,C) = (1.5,0.08,0.16) and $\alpha$ = 1.25. (a) the bifurcation diagram, and (b) the variation of two largest Lyapunov exponents.}
\end{figure*}

\section{Conclusion}
The present work studies the advection of finite sized passive scalar particles in an incompressible three-dimensional flow for cases where the particle density  differs from the fluid density. The motion of the advected particles is represented in Lagrangian description by an embedding map with the volume preserving Arnold-Beltrami-Childress (ABC) map as the base map. The resulting embedded ABC map is invertible and dissipative with two sets of parameters namely $(A,B,C)$ which belong to the base map and the mass ratio and  dissipation parameters $(\alpha,\gamma)$ which  enter due to the embedding. The two-action case of the map has been studied in both the regimes - aerosol and bubble, depending upon whether the mass ratio parameter, $\alpha<1$ or $\alpha>1$.

In the two-action case, the phase diagram for the system in the $(\alpha-\gamma)$ parameter space shows rich structures with complex dynamics.  Three type of dynamical behaviors - periodic orbits, chaotic and hyperchaotic regions are found to be present. The bubble regime is mostly hyperchaotic, but the aerosol regime also contains a tongue of hyperchaoticity at low values of dissipation.  Crisis induced intermittency in this region, and   power law behavior for the characteristic times between bursts is seen  in the neighborhood of the crisis. Unstable dimension variability is also seen in this neighborhood. The bubble regime also shows the existence of multiple co-existing attractors, and a riddled basin of attraction. The observed riddling may be a consequence of a bubbling bifurcation, similar to that seen in the bailout embeddings of Hamiltonian dynamical systems \cite{Julyan}. This needs to be explored further, as well as the consequences of the riddling for clustering and the preferential concentration of particles, also need to be examined. We hope to do this in future work. 

The aerosol regime of the two action case exhibits a bifurcation diagram with rich structure. An interior crisis is seen in the system, and in fact, a number of windows of crisis are seen. A two ring attractor is seen in the post crisis setting, with trajectories which hop between the rings with period two. The transient to the two ring attractor shows scaling behavior. 
Thus, our toy model shows a variety of phenomena which could lead to consequences for transport and  pattern formation.  Clustering and concentration phenomena and their dependence on initial conditions have been insufficiently explored so far, even in model contexts. The present system constitutes an excellent toy model for exploring such phenomena. We hope to explore these phenomena as well as the drift and diffusive properties of this system in future work. 

A preliminary study has also been carried out for the one action case of the $ABC$ map, using the parameter values $A=1.5, B=0.08, C=0.16$. Here, in the
aerosol regime with $\alpha=0.2$, the
largest LE is negative for most of the range $0\leq\gamma\leq4$.
Here,  periodic attractors with even periods  were seen. The
bubble regime with $\alpha=1.25$ showed fully hyperchaotic behavior where the two largest LEs were  always positive in the range
$0\leq\gamma\leq4$, and some of the phase space plots showed patched structure. However, a detailed study of the one action case remains to be carried
out. It is expected that the presence of invariant surfaces in this case will lend interesting features to the drift and diffusion properties.

It would also be interesting to connect the present method to other approaches, e.g. that of Lagrangian Coherent Structure analysis \cite{Yang,Shadden}.  In addition to the fact that the system studied here exhibits dynamical phenomena which are of theoretical interest, our observations may have implications in a variety of application contexts such as the dispersion of pollutants in the atmosphere and oceans, catalytic chemical reactions, coagulation of material particles in flows, plankton population in ocean etc.  We also note that the transport properties and physical behavior of impurities in three dimensions are little understood at present. Finally, given that the ABC flow and map constitute models for magnetohydrodynamics, their dynamics and transport properties may have implications for magnetohydrodynamic properties like the dynamo effect. We hope to analyze these aspects, and their consequences for applications in future work.

\bibliographystyle{plainnat}

\begin{thebibliography}{widest-entry}
\bibitem{Integrability} 
Integrability may be seen when two of parameters vanish, say $B$ and $C$ and the motion will be confined to surfaces of constant  $z$ and the quantity $\sqrt {(x\prime -x)^2 + (y\prime -y)^2}$ is a constant which is equal to the parameter $A$. For further details of the quasi-integrability of the ABC map, see\cite{Feingold}. In case of the ABC flow, it is straightforward to show that the flow becomes integrable when one of the three parameters vanishes\cite{Dombre}.
\bibitem{Bifur}
A similar diagram is seen if $\frac{y}{2\pi}$ and $\frac{z}{2\pi}$ variables are plotted versus $\gamma$.
\bibitem{RCD}
The reverse cumulative distribution plotted here is the cumulative distribution in reverse order. By definition, the largest value in the cumulative distribution appears on the left of the plot and corresponds to the highest point in the curve. The lowest point on the curve (to the right of the plot) indicates the frequency of the largest $x$-value observed.
\bibitem{FN}
In Ref. \cite{Nirmal}, trajectories have been defined to be mixing if they access $99 \%$ of the phase space.
\end{thebibliography}

\end{document}